\documentclass[10pt,twocolumn,letterpaper]{article}

\usepackage{wacv}
\usepackage{times}
\usepackage{epsfig}
\usepackage{graphicx}
\usepackage{amsmath}
\usepackage{amssymb}
\usepackage{booktabs}
\usepackage[subrefformat=parens]{subcaption}
\usepackage{url}
\usepackage{algorithm}
\usepackage{algpseudocode}
\usepackage[accsupp]{axessibility}

\def\x{{\mathbf x}}
\def\y{{\mathbf y}}

\def\thetabold{\boldsymbol{\theta}}
\def\phibold{\boldsymbol{\phi}}

\newcommand{\argmin}{\mathop{\rm arg~min}\limits}

\def\wacvPaperID{1826} 
\wacvalgorithmstrack   

\wacvfinalcopy 

\ifwacvfinal
\usepackage[breaklinks=true,bookmarks=false]{hyperref}
\else
\usepackage[pagebackref=true,breaklinks=true,colorlinks,bookmarks=false]{hyperref}
\fi

\begin{document}

\title{Universal Deep Image Compression\\ via Content-Adaptive Optimization with Adapters}

\author{Koki Tsubota$^{1}$~~~~~~Hiroaki Akutsu$^{2}$~~~~~~Kiyoharu Aizawa$^{1}$\\
$^{1}$The University of Tokyo~~~~~~~~$^{2}$Hitachi, Ltd.\\
{\tt\small \{tsubota,aizawa\}@hal.t.u-tokyo.ac.jp, hiroaki.akutsu.cs.@hitachi.com}
}

\maketitle
\thispagestyle{empty}

\begin{abstract}
Deep image compression performs better than conventional codecs, such as JPEG, on natural images.
However, deep image compression is learning-based and encounters a problem: the compression performance deteriorates significantly for out-of-domain images.
In this study, we highlight this problem and address a novel task: universal deep image compression.
This task aims to compress images belonging to arbitrary domains, such as natural images, line drawings, and comics.
To address this problem, we propose a content-adaptive optimization framework; this framework uses a pre-trained compression model and adapts the model to a target image during compression.
Adapters are inserted into the decoder of the model.
For each input image, our framework optimizes the latent representation extracted by the encoder and the adapter parameters in terms of rate-distortion.
The adapter parameters are additionally transmitted per image.
For the experiments, a benchmark dataset containing uncompressed images of four domains (natural images, line drawings, comics, and vector arts) is constructed and the proposed universal deep compression is evaluated.
Finally, the proposed model is compared with non-adaptive and existing adaptive compression models.
The comparison reveals that the proposed model outperforms these.
The code and dataset are publicly available at \url{https://github.com/kktsubota/universal-dic}.
\end{abstract}

\section{Introduction}
Image compression is a fundamental technology for reducing the costs of storage and network transmission.
Compressed images are ubiquitous---digital cameras and smartphones compress images.
The common compression standard is JPEG~\cite{JPEG}, whereas JPEG2000~\cite{JPEG2000}, BPG~\cite{BPG}, and VVC~\cite{VVC} are more recent standard-based compression.
Deep image compression is the image compression technique based on neural networks.
Recent studies have demonstrated that deep image compression exhibits higher performance than conventional codecs on natural images~\cite{guo2021causal,ZhuICLR22,ZouCVPR22}.

\begin{figure}[t]
\begin{minipage}{0.48\hsize}
    \centering
    \includegraphics[width=\hsize]{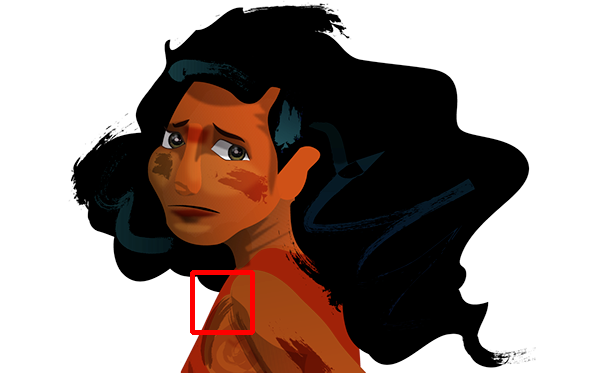}
    \captionsetup{justification=centering}
    \subcaption*{Input\\(PSNR/BPP)}
  \end{minipage}
  \begin{minipage}{0.48\hsize}
    \centering
    \includegraphics[width=\hsize]{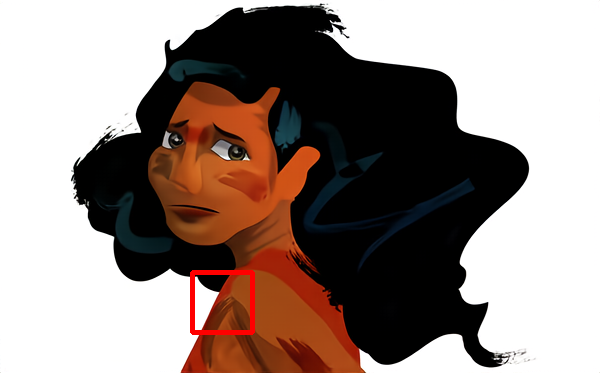}
    \captionsetup{justification=centering}
    \subcaption*{Ours\\(\textbf{39.2}/0.112)}
  \end{minipage}\\

  \begin{minipage}{0.48\hsize}
    \centering
    \includegraphics[width=\hsize]{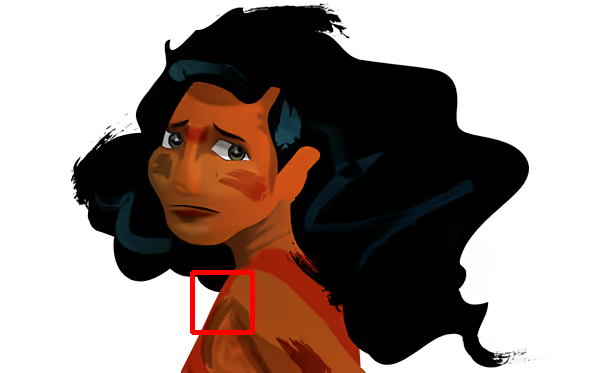}
    \captionsetup{justification=centering}
    \subcaption*{VVC\\(38.5/0.108)}
  \end{minipage}
  \begin{minipage}{0.48\hsize}
    \centering
    \includegraphics[width=\hsize]{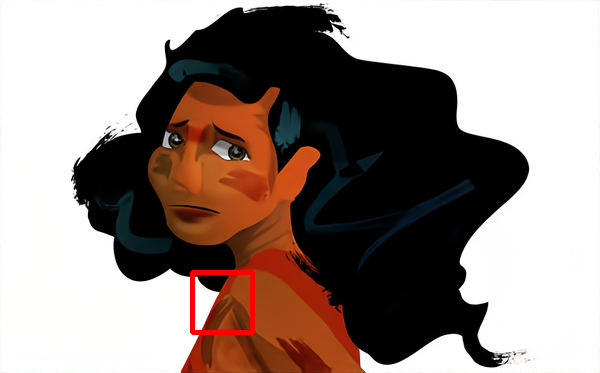}
    \captionsetup{justification=centering}
    \subcaption*{WACNN~\cite{ZouCVPR22}\\(38.3/0.109)}
  \end{minipage}

  \begin{minipage}{0.24\hsize}
    \centering
    \setlength{\fboxrule}{1pt}
    \setlength{\fboxsep}{0pt}
    \fcolorbox{red}{white}{\includegraphics[width=\hsize]{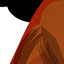}}
    \captionsetup{justification=centering}
    \subcaption*{Input}
  \end{minipage}
  \begin{minipage}{0.24\hsize}
    \centering
    \setlength{\fboxrule}{1pt}
    \setlength{\fboxsep}{0pt}
    \fcolorbox{red}{white}{\includegraphics[width=\hsize]{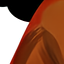}}
    \captionsetup{justification=centering}
    \subcaption*{Ours}
  \end{minipage}
  \begin{minipage}{0.24\hsize}
    \centering
    \setlength{\fboxrule}{1pt}
    \setlength{\fboxsep}{0pt}
    \fcolorbox{red}{white}{\includegraphics[width=\hsize]{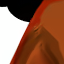}}
    \captionsetup{justification=centering}
    \subcaption*{VVC}
  \end{minipage}
  \begin{minipage}{0.24\hsize}
    \centering
    \setlength{\fboxrule}{1pt}
    \setlength{\fboxsep}{0pt}
    \fcolorbox{red}{white}{\includegraphics[width=\hsize]{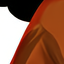}}
    \captionsetup{justification=centering}
    \subcaption*{WACNN~\cite{ZouCVPR22}}
  \end{minipage}
  \caption{Examples of compression results on a comic image in the BAM dataset~\cite{BAM}. General deep image compression (WACNN~\cite{ZouCVPR22}) performs superior to the state-of-the-art conventional codec (VVC~\cite{VVC}) on natural images. However, its performance deteriorates on out-of-domain images. By addressing this problem, our framework exhibits superior performance to VVC on out-of-domain images. In this figure, ours can reconstruct the brush texture in dark brown with relatively high fidelity.}
  \label{fig:sample}
  \vspace{-8pt}
\end{figure}

However, deep image compression is learning-based.
Therefore, we encounter the problem of performance degradation in compressing out-of-domain images.
General compression models pre-trained only on natural images exhibit relatively low performance on images in other domains, as shown in Fig.~\ref{fig:sample}.
To investigate this problem, we address a novel deep image compression task, which we name universal deep image compression.
The objective of universal deep image compression is to compress images from arbitrary domains, such as line drawings and comics, as well as natural images.

We propose a content-adaptive optimization framework to address the problem of compressing out-of-domain images.
This framework adapts the pre-trained compression model to each target image and addresses domain shifts between pre-training and testing.
Our framework is efficient owing to a small number of parameters needed for the per-image adaptation during testing.

Our framework has two advantages over previous approaches studied in content-adaptive compression~\cite{conf/cvpr/CamposMDS19,conf/cvprw/LamZACLAH19,conf/mm/LamZCLH20,rozendaal2021overfitting,ZouMMSP20,ZouAdaptISM21}: the flexibility of the base network architecture and the efficiency in terms of rate-distortion.
In content-adaptive compression, certain studies adapted the compression model during testing by refining the latent representation extracted by the encoder~\cite{conf/cvpr/CamposMDS19,conf/nips/Yang20}.
Other studies additionally updated the parameters in the decoder and transmitted these~\cite{conf/cvprw/LamZACLAH19,conf/mm/LamZCLH20,rozendaal2021overfitting,ZouMMSP20,ZouAdaptISM21}.
However, the state-of-the-art latent refinement method~\cite{conf/nips/Yang20} has restrictions on pre-trained compression models: it assumes that the hyper latent representation follows a Gaussian distribution to perform the bit-back coding~\cite{BitbackCodingHintonC93,BitbackCodingWallace90}.
Moreover, previous approaches for updating the parameters in the decoder update an excessive number of parameters for an individual image~\cite{conf/mm/LamZCLH20,rozendaal2021overfitting}, insert and train ad-hoc layers~\cite{ZouAdaptISM21}, or optimize parameters only in terms of distortion~\cite{conf/mm/LamZCLH20,ZouAdaptISM21}.

In our framework, we refine the latent representation by the simplified approach of the state-of-the-art refinement method.
We omit the process of bit-back coding and only optimize the latent representation in terms of rate-distortion with gradient descent.
Therefore, our latent refinement is efficient and flexible to the base network architecture.

To update the decoder, we insert adapters into the decoder and train these.
Adapters are small modules with a small number of parameters and have been successful in parameter-efficient transfer learning~\cite{conf/icml/HoulsbyGJMLGAG19,LiCVPR22TaskSpecificAdapter,conf/nips/RebuffiBV17,Sung_2022_CVPR}.
Using adapters, we can improve the compression performance by updating a relatively small number of parameters.
Moreover, we optimize adapters in terms of rate-distortion with gradient descent.
Therefore, our decoder update is efficient in terms of rate-distortion.

To evaluate our framework, we construct a benchmark dataset that comprises four domains: natural images, line drawings, comics, and vector arts.
We sample natural images from the Kodak dataset~\cite{Kodak} and images in the other three domains from the BAM dataset~\cite{BAM}.
We use one of the state-of-the-art compression models (window attention-based convolutional neural networks (WACNN)~\cite{ZouCVPR22})
for the baseline, and modify it by inserting adapters and optimizing the latent representation and the adapter parameters.
We pre-train the model on a natural image dataset and evaluate its performance on in-domain and out-of-domain images.

The main contributions of this study are as follows
\begin{itemize}
  \item We address universal deep image compression. To our knowledge, this is the first work that addresses the deep compression of images in arbitrary domains, such as line drawings and comics.
  \item We propose a content-adaptive optimization framework, wherein we adapt a pre-trained model to each target image. Our framework refines the latent representation by a simplified approach of the state-of-the-art method. We then train adapters inserted into the decoder via optimization in terms of rate-distortion. The adapter parameters are additionally transmitted.
  \item We demonstrate experimentally that the proposed method is effective and significantly outperforms the state-of-the-art conventional codec on the four domains.
\end{itemize}

\begin{figure*}[t]
  \centering
  \includegraphics[width=\hsize]{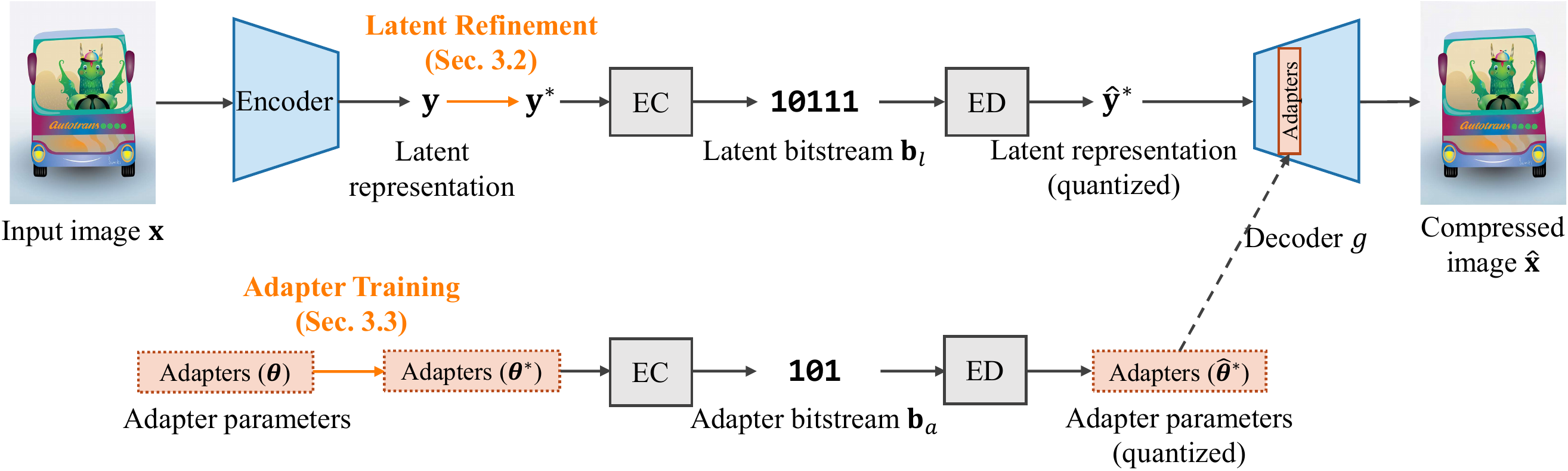}
  \caption{Outline of the proposed method. First, we refine the latent representation; subsequently, we train the adapters.}
  \label{fig:method}
  \vspace{-8pt}
\end{figure*}

\section{Related Work}
\subsection{Deep Image Compression}
Deep image compression achieves image compression by optimizing the modules in an end-to-end manner~\cite{conf/iclr/BalleLS17}.
To obtain compressed images with less distortion, numerous studies have worked toward improving the modules such as the encoders and decoders~\cite{conf/cvpr/Cheng2020,ZhuICLR22} and the entropy models~\cite{conf/cvpr/MentzerATTG18,conf/nips/MinnenBT18,conf/icip/MinnenS20}.
Some studies worked on image compression for human perception instead of distortion~\cite{conf/icml/BlauM19,MentzerHific,Patel_2021_WACV,conf/icml/RippelB17}.
Other studies worked on achieving variable rate compression~\cite{conf/iccv/ChoiEL19}.

Deep image compression outperforms conventional codecs on natural images~\cite{guo2021causal,ZhuICLR22,ZouCVPR22}.
However, these studies trained only on natural images, such as CLIC~\cite{CLIC20} and ImageNet~\cite{ImageNet}, and evaluated only on natural images, such as Kodak~\cite{Kodak}, CLIC~\cite{CLIC20}, Tecnick~\cite{Tecnick}, and DIV2K~\cite{Div2k}.
Hence, their compression performance when applied to other domains, such as line drawings and comics, remains uncertain.

In contrast, Kim \etal~\cite{KimPruning20} worked on lightweight and fast decoding in deep image compression.
They evaluated their proposed method on both natural and cartoon images.
However, their approach needs to prepare a dataset for training on cartoon images, unlike a content-adaptive optimization framework that requires no pre-training per domain.
Moreover, the number of evaluation domains is limited.
In this study, we study the compression of images in various domains, such as line drawings, comics, and natural images.

\subsection{Content-Adaptive Compression}
Content-adaptive compression achieves high performance by adapting the compression model for each target image.
In deep image compression, content-adaptive compression is achieved by the per-image refinement of the latent representation obtained by the encoder~\cite{conf/cvpr/CamposMDS19,conf/nips/Yang20}.
The latent bitstream can be obtained by compressing the refined latent representation.
In addition to the latent refinement, Zou \etal~\cite{ZouMMSP20} updated the parameters in the decoder per image.
The model bitstream can be obtained by compressing the updated parameters.

Updating and compressing the parameters in the decoder has been studied primarily for compression of multiple images, post-processing of video compression, and deep video compression.
Rozendaal \etal~\cite{rozendaal2021overfitting} updated all the parameters in the decoder and entropy model.
They optimized the parameters in terms of rate-distortion and finally compressed these by entropy coding.
However, although they addressed multiple images for the adaptation, this approach requires a relatively large number of bits for compressing an individual image.

Other studies updated the limited number of adapting parameters.
Zou \etal~\cite{ZouMMSP20} addressed deep image compression and updated only the biases of convolution layers in the decoder.
Lam \etal~\cite{conf/mm/LamZCLH20} addressed the post-processing of compressed videos and updated only the biases of convolutional layers in the post-processing network.
Zou \etal~\cite{ZouAdaptISM21} addressed deep video compression. They inserted overfittable multiplicative parameters (which multiply the output of convolutional layers) and updated these for intra-frame coding.
However, updating parameters are selected in an ad-hoc manner and are optimized only in terms of distortion.
Unlike these previous approaches, we introduce adapters and optimize these in terms of rate-distortion.
Adapters exhibit superior performance in parameter-efficient transfer learning.

\subsection{Parameter-Efficient Transfer Learning}
Parameter-efficient transfer learning aims to adapt a model pre-trained on a large-scale dataset per task, with reducing the number of adapting parameters.
Unlike a general adaptation approach that fine-tunes all the parameters of a pre-trained model, most of the model parameters are fixed.
Therefore, we can reduce the cost of transmission of parameters and preserve the knowledge in pre-training.

This task was first studied for obtaining a universal representation of multiple domains in computer vision~\cite{conf/nips/RebuffiBV17,conf/cvpr/RebuffiBV18}.
Recently, motivated by the emergence of a big pre-trained transformer model~\cite{Transformer}, such as BERT~\cite{NLP_BERT} and T5~\cite{NLP_T5}, this task has been studied primarily for training efficiently in natural image processing~\cite{ZakenACL22_BitFit,conf/icml/HoulsbyGJMLGAG19}.

We can classify the algorithms for parameter-efficient transfer learning into the following two types.
(1) adapting newly added parameters~\cite{HeICLR22,MahabadiNeurIPS21_Compacter,LiCVPR22TaskSpecificAdapter,conf/nips/RebuffiBV17,conf/cvpr/RebuffiBV18} and (2) adapting part of the parameters in the model~\cite{ZakenACL22_BitFit,GuoACL21_DiffPruning,conf/iclr/MudrakartaSZH19}.
The first approach introduces an adaptation module, such as adapters~\cite{conf/nips/RebuffiBV17,conf/cvpr/RebuffiBV18}, compacters~\cite{MahabadiNeurIPS21_Compacter}, and hyperformers~\cite{KarimiACL21_Hyperformer}. The second approach adapts normalization and lightweight layers~\cite{conf/iclr/MudrakartaSZH19}, biases of layers in the model~\cite{ZakenACL22_BitFit}, and sparse difference in model weights~\cite{GuoACL21_DiffPruning}.

Among these approaches, adapters are widely used~\cite{HeICLR22,conf/icml/HoulsbyGJMLGAG19,LiCVPR22TaskSpecificAdapter,conf/nips/RebuffiBV17,conf/cvpr/RebuffiBV18,Sung_2022_CVPR}.
Adapters are modules with a small number of parameters.
These are implemented as matrix multiplication~\cite{conf/cvpr/RebuffiBV18}, decomposition of a matrix~\cite{conf/cvpr/RebuffiBV18}, multiplication of two matrices with activation~\cite{conf/icml/HoulsbyGJMLGAG19}, or channel-wise scaling~\cite{LiCVPR22TaskSpecificAdapter}.
Motivated by the success of adapters in this task, we used adapters in our framework.

\section{Method}
\begin{figure}[t]
  \centering
  \includegraphics[width=\hsize]{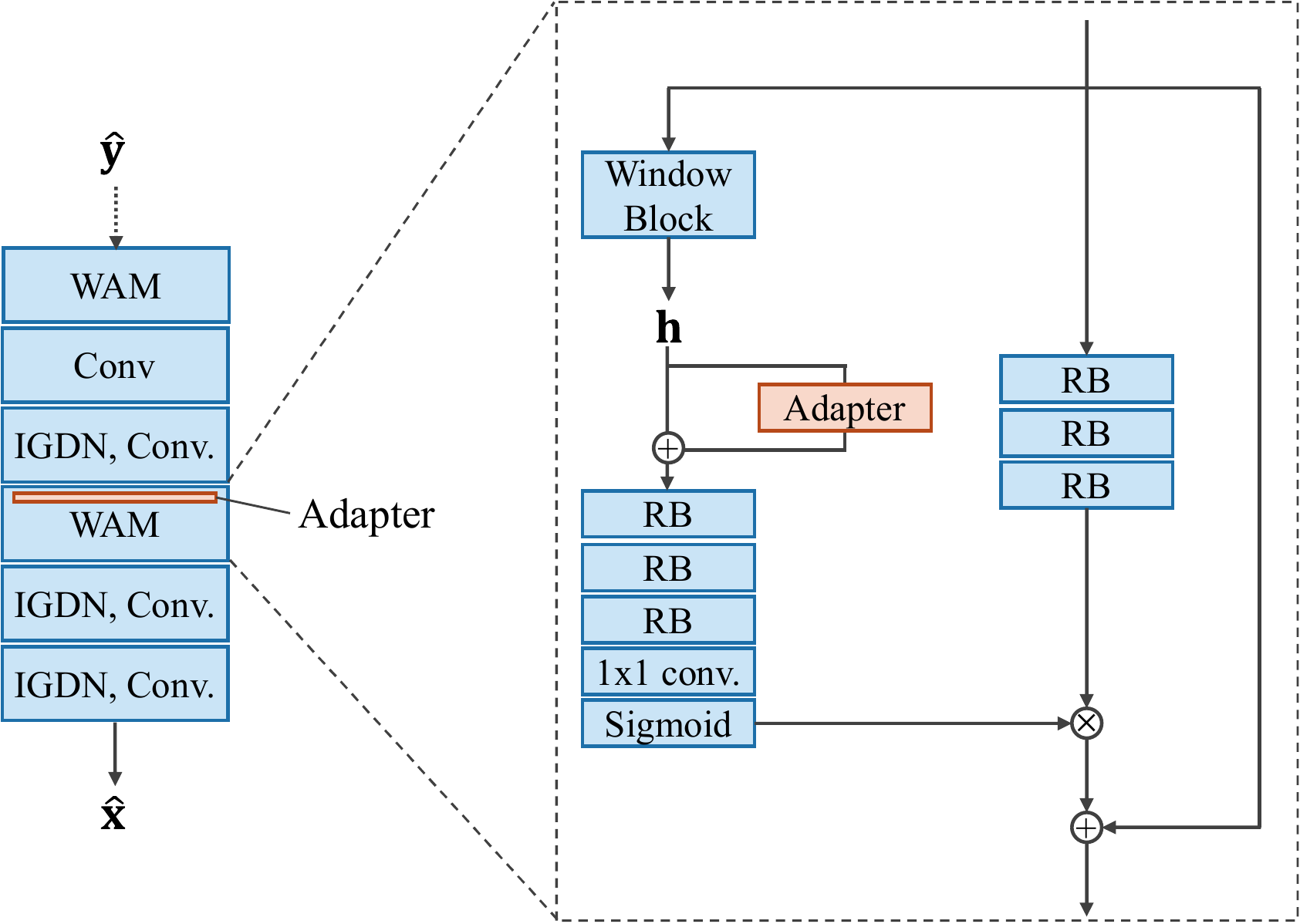}
  \caption{Network architecture of the main decoder with adapters. WAM, Conv., IGDN, and RB denote a window-attention module~\cite{ZouCVPR22}, a convolutional layer, an inverse generalizable divisible normalization layer~\cite{Balle/iclr/GDN16}, and a residual block~\cite{conf/cvpr/HeZRS16}, respectively.}
  \label{fig:adapter}
  \vspace{-8pt}
\end{figure}

The objective of universal deep image compression is to compress images from arbitrary domains.
To achieve this objective, we proposed a content-adaptive optimization framework.
Given a pre-trained compression model and an uncompressed image, the model is adapted to each target image during testing.

The outline of the proposed method is shown in Fig.~\ref{fig:method}.
In encoding, first, latent representation is extracted by the encoder and adapters are inserted into the decoder.
Subsequently, the latent representation is refined via optimization in terms of rate-distortion.
Next, the adapters are trained by optimizing in terms of rate-distortion.
Finally, the latent representation and the parameters of the adapters are encoded by entropy coding, and the latent and adapter bitstreams are transmitted.
In decoding, the transmitted bitstreams are decoded by entropy decoding, and finally, the compressed image is obtained.

We used WACNN~\cite{ZouCVPR22}, which is a state-of-the-art architecture, as the base network architecture.
Note that our framework can be applied to other network architectures.
WACNN has a hyper-prior architecture~\cite{conf/iclr/BalleMSHJ18}.
It transmits the hyper latent representation as well as the latent representation.
In our explanation, we considered these two representations in a unified manner and called these as the latent representation.

Hereafter, we explain the details of our framework.
Our framework comprises the following three technical components.
The first is the insertion of the adapters into the decoder.
The second is the refinement of the latent representation extracted from the target image.
The third is the training of the adapters.
We describe the details of each component in the next subsections.

Let us define the characters for the explanation.
Let $\x$ be the input image, $\hat{\x}$ be the compressed image, $\y$ be the latent representation, $\hat{\y}$ be the quantized latent representation, $\y^{*}$ be the refined latent representation, $q$ be the quantizer, $g$ be the decoder with adapters, $\mathbf{b}_l$ be the latent bitstream, $\mathbf{b}_a$ be the adapter bitstream, $\phibold$ be the pre-trained parameters for the decoder except for adapters, $\thetabold$ be the parameters of the adapters, and $\thetabold^{*}$ be the updated parameters of the adapters.

\subsection{Insertion of Adapters}
We empirically determined good insertion positions of the adapters and inserted an adapter into the window attention module (WAM)~\cite{ZouCVPR22} on the second side of the main decoder.
The network architecture of the main decoder with adapters is shown in Fig.~\ref{fig:adapter}.

We implemented the adapter by matrix decomposition presented in \cite{conf/cvpr/RebuffiBV18}.
This architecture is simple albeit effective, as shown in \cite{LiCVPR22TaskSpecificAdapter}.
Let the input of the adapter be $\mathbf{h} \in \mathbb{R}^{C \times H \times W}$ and the adapter be $r: \mathbb{R}^{C\times H \times W} \to \mathbb{R}^{C\times H \times W}$.
The operation of the adapter is written as
\begin{align}
    r(\mathbf{h}; \thetabold) = \mathbf{AB}^{\top}\mathbf{h},
\end{align}
where $\mathbf{A}, \mathbf{B} \in \mathbb{R}^{C\times M}$ are learnable parameters of the adapter and $\thetabold = [\mathbf{A}, \mathbf{B}]$.
Thus, the number of adapter parameters is $2MC$.

The number of adapter parameters is much smaller than the number of model parameters.
For WACNN~\cite{ZouCVPR22}, the number of model parameters is $6.50 \times 10^{7}$ and $C = 192$.
Thus, when $M = 2$, the number of adapter parameters is 768.
This is 0.0012\% to the model parameters.

\subsection{Latent Refinement}
\begin{figure}[t]
  \centering
  \includegraphics[width=\hsize]{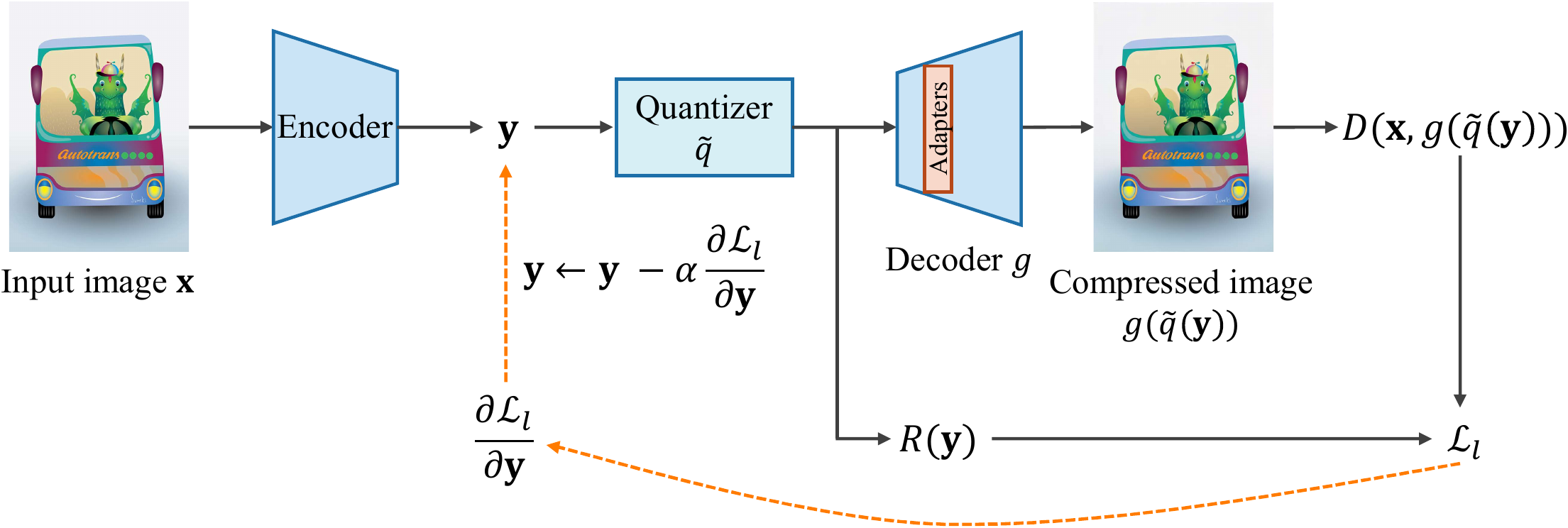}
  \caption{Outline of latent refinement.}
  \label{fig:latent_adaptation}
  \vspace{-8pt}
\end{figure}

\begin{figure}[t]
  \centering
  \includegraphics[width=\hsize]{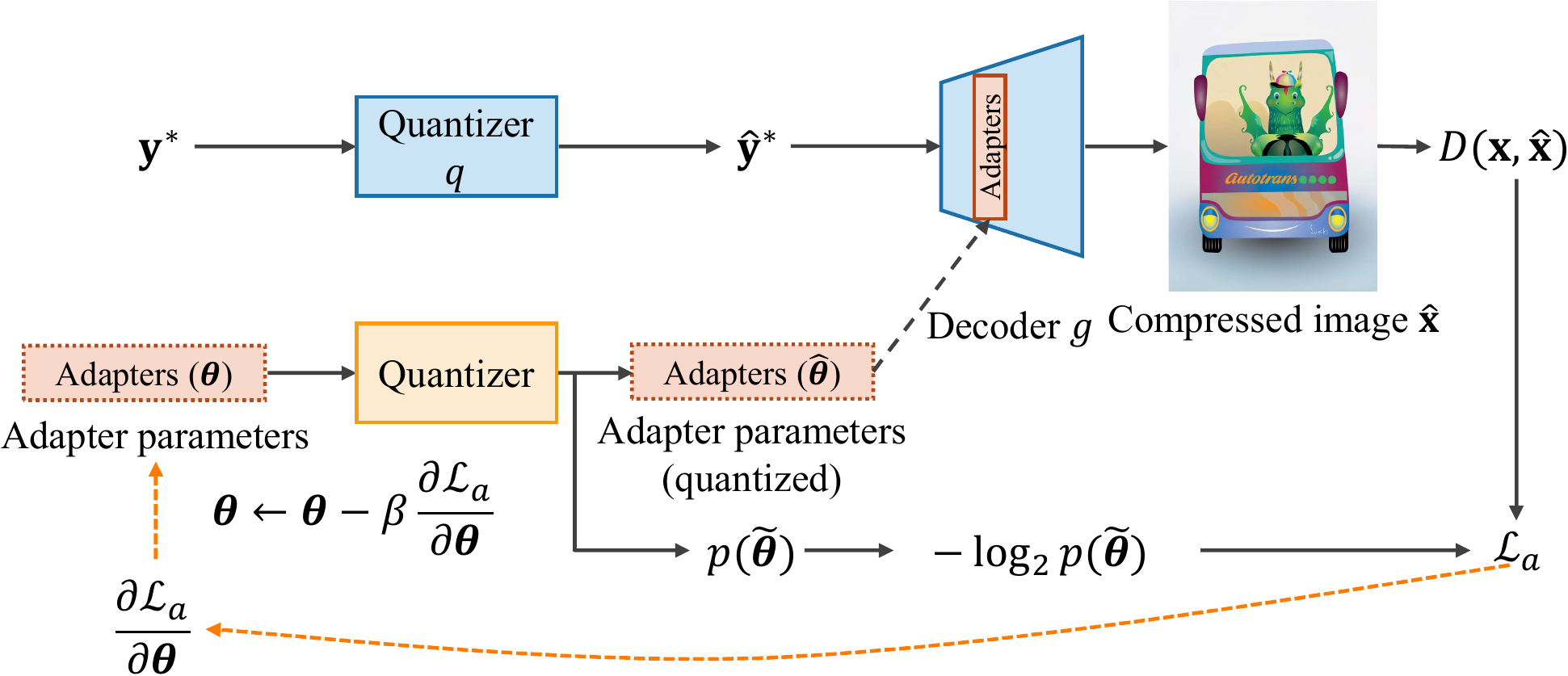}
  \caption{Outline of adapter training.}
  \label{fig:adapter_training}
  \vspace{-8pt}
\end{figure}

The outline of the latent refinement is illustrated in Fig.~\ref{fig:latent_adaptation}.
We optimized the latent representation $\y$ with gradient descent via the simplified approach of Yang \etal~\cite{conf/nips/Yang20}.
The loss function is written as
\begin{align}\label{eq:loss_latent}
    \mathcal{L}_l (\y) = R (\tilde{q}(\y)) + \lambda D \left( g \left(\tilde{q}(\y); \phibold, \thetabold\right), \x \right),
\end{align}
where $R$ is bitrate, $D$ is the distortion, $\lambda \in \mathbb{R}$ is the hyper-parameter for adjusting the trade-off of rate-distortion, and $\tilde{q}$ is the uniform quantization with the approximation of stochastic Gumbel annneling~\cite{conf/nips/Yang20}.
In this optimization, we did not train the adapters and fix the parameters to zero.
We omitted bit-back coding in Yang \etal~\cite{conf/nips/Yang20} because it required modification of the pre-trained network architecture.

We obtained the refined latent representation $\y^{*}$, which is given as follows.
\begin{equation}
    \y^{*} = \argmin_{\y} \mathcal{L}_l (\y).
\end{equation}
Finally, we quantized $\y^{*}$ and encoded $\hat{\y}^{*}$ by entropy coding to obtain $\mathbf{b}_l$.
Note that this latent refinement can be completed using a local decoder at the transmitter.

\subsection{Adapter Training}
The outline of adapter training is shown in Fig.~\ref{fig:adapter_training}.
We optimized the parameters of the adapters as the optimization of the latent representation.
Let $w$ be the quantization interval and $\hat{\thetabold}$ be the quantized adapter parameters.
We quantized $\thetabold$ with approximation and optimized $\thetabold$ in terms of rate-distortion.
We used the mixed quantization approach~\cite{conf/iclr/LeeCB19}.
That is, we uniformly quantized $\thetabold$ with a straight-through estimator~\cite{conf/nips/HubaraCSEB16} for the decoder and add uniform noise $U(-w/2, w/2)$ to $\thetabold$ for the entropy model.
Let $\tilde{\thetabold}$ be the adapter parameters added uniform noise.

The loss function for optimizing $\thetabold$ is written as
\begin{equation}\label{eq:loss_adapter}
    \mathcal{L}_a (\thetabold) = - \log_2 p (\tilde{{\thetabold}}) + \lambda D \left( g \left(\hat{\y}; \phibold, \hat{\thetabold} \right), \x \right),
\end{equation}
where $p$ is the entropy model of $\thetabold$.
The first term is the loss function for the bitrate and computes the entropy of $\hat{\thetabold}$.
We used a logistic distribution with the scale of $s \in \mathbb{R}$ as $p$.

After this optimization, we obtained the updated parameters of the adapters $\thetabold^{*}$, which are given as follows.
\begin{equation}
    \thetabold^{*} = \argmin_{\thetabold} \mathcal{L}_a (\thetabold).
\end{equation}
Finally, we quantized $\thetabold^{*}$ and encoded $\hat{\thetabold}^{*}$ by entropy coding to obtain $\mathbf{b}_a$.

\section{Experiment}
\subsection{Experimental Setup}
We constructed a benchmark dataset containing uncompressed images of four domains: natural images, comics, line drawings, and vector arts.
We collected natural images from the Kodak dataset~\cite{Kodak} and the images in the other domains from the BAM dataset~\cite{BAM}.
The Kodak dataset consists of 24 natural images.
We used all the images in the Kodak dataset.
The BAM dataset consists of artistic images other than natural images.
We sampled 100 images from the BAM dataset per domain that were not degraded by JPEG compression.
We considered images labeled with ``pen-ink'', ``comic'', and ``vectorart'' as line drawings, comics, and vector arts, respectively.
The statistics of the constructed dataset are presented in Table~\ref{tbl:bam}.

We used WACNN~\cite{ZouCVPR22} implemented with CompressAI~\cite{compressai} as the base compression model.
We pre-trained six models with $\mathcal{L} = R + \lambda D$ by setting $\lambda$ to 0.0018, 0.0035, 0.0067, 0.013, 0.025, and 0.0483.
The pre-training data comprised 300,000 natural images sampled randomly from OpenImages~\cite{OpenImages}.
Thus, the results for the natural images indicate in-domain performance, whereas those for the other three domains indicate out-of-domain performance.

With regard to the hyper-parameters, we set $w = 0.06$, $s = 0.05$, and $M = 2$ in the proposed method.
We set $\lambda$ to an equal value in pre-training.
We use mean squared error as distortion $D$.

\begin{table}[t]
  \centering
  \caption{Evaluation dataset.}
  \label{tbl:bam}
  \begin{tabular}{lcc}
    \toprule
    Domain & Test data & Average Resolution\\
    \midrule
    Natural image & 24 & $576\times704$\\
    Comic & 100 & $606\times587$\\
    Line drawing & 100 & $584\times577$\\
    Vector art & 100 & $554\times580$\\
    \bottomrule
  \end{tabular}
  \vspace{-8pt}
\end{table}

\textbf{Implementation Details.}
In pre-training, we used the Adam optimizer~\cite{kingma2015adam} for up to 100 epochs with a batch size of 16.
We set the learning rate to $10^{-3}$ for the first 78 epochs, $10^{-4}$ for the following 20 epochs, and $10^{-5}$ for the final two epochs.

In adaptation, we used the Adam optimizer for up to 2,000 iterations for the latent refinement and 500 iterations for the adapter training.
We set the learning rate to $10^{-3}$ for the first 1,600 iterations and $10^{-4}$ for the final 400 iterations for the latent refinement.
We set the learning rate to $10^{-3}$ for the first 400 iterations and $10^{-4}$ for the final 100 iterations for the adapter training.
The $\thetabold_a$ is initialized with the Gaussian noise of $\mathcal{N}(0, 0.02^2)$.
For further details, please refer to our publicly available source code.

\begin{figure*}[t]
  \centering
  \begin{minipage}{0.48\hsize}
    \centering
    \includegraphics[width=0.95\hsize]{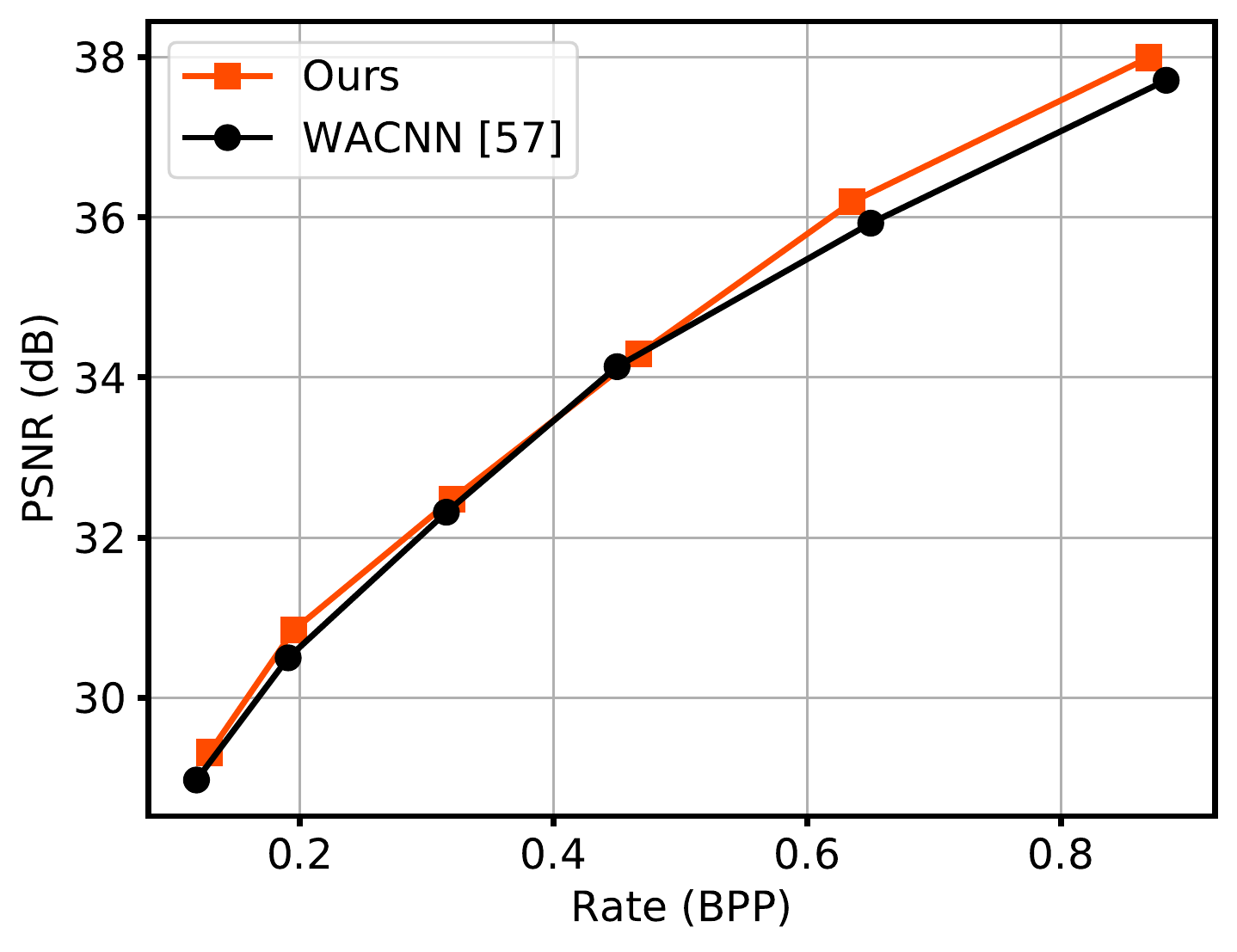}
    \vskip-5pt
    \subcaption*{Natural image}
  \end{minipage}
  \begin{minipage}{0.48\hsize}
    \centering
    \includegraphics[width=0.95\hsize]{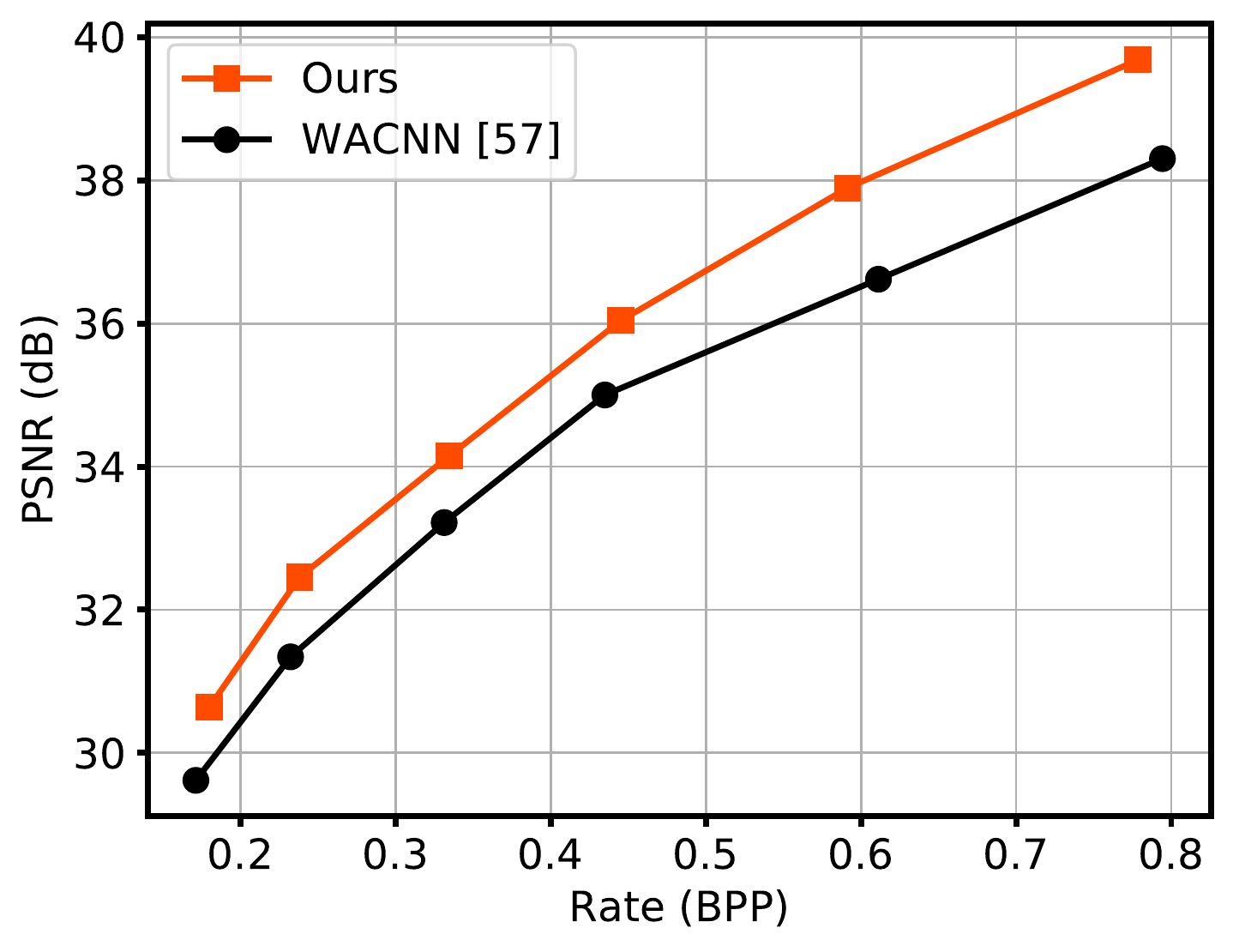}
    \vskip-5pt
    \subcaption*{Comic}
  \end{minipage}

  \begin{minipage}{0.48\hsize}
    \centering
    \includegraphics[width=0.95\hsize]{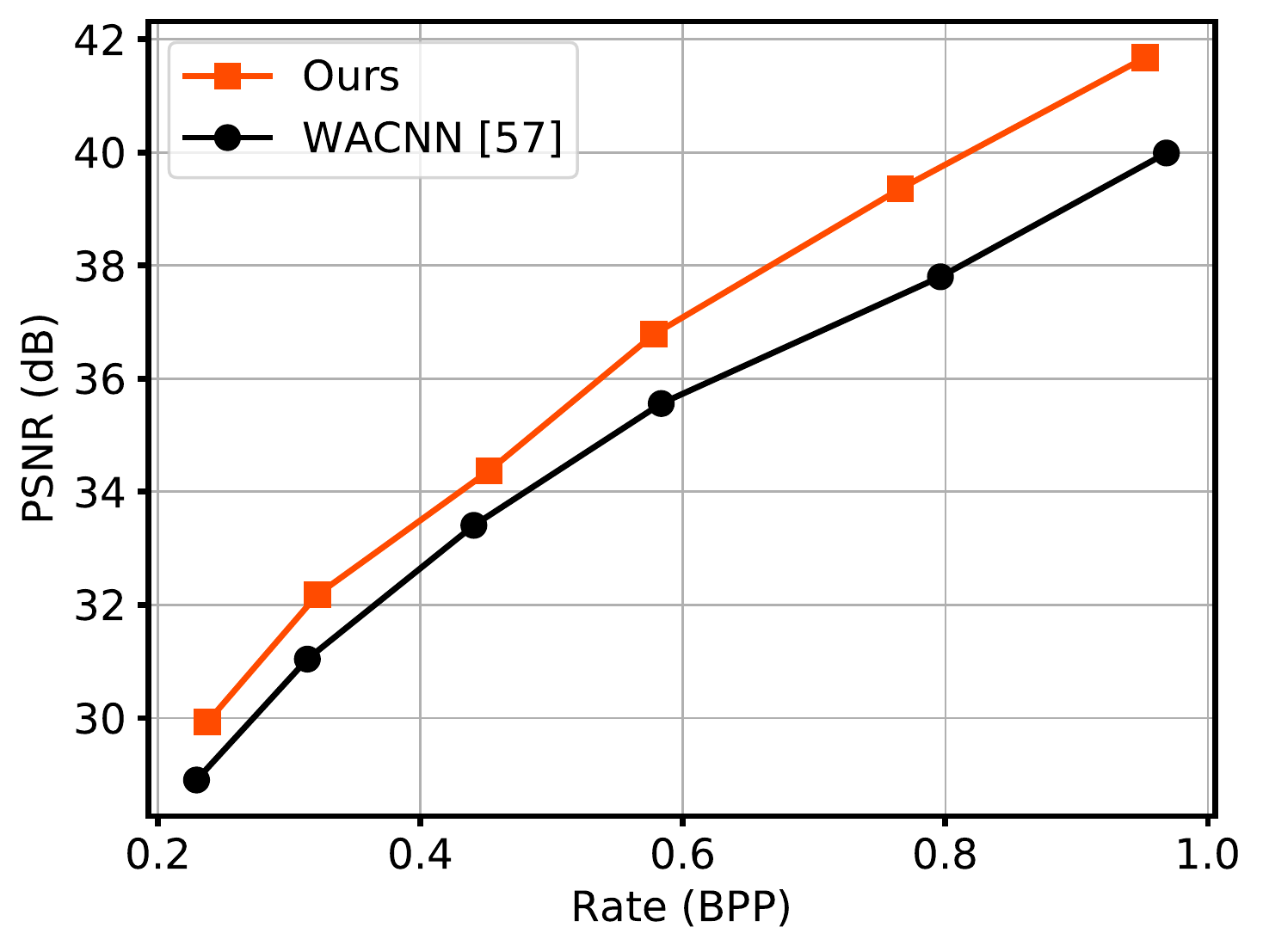}
    \vskip-5pt
    \subcaption*{Line drawing}
  \end{minipage}
  \begin{minipage}{0.48\hsize}
    \centering
    \includegraphics[width=0.95\hsize]{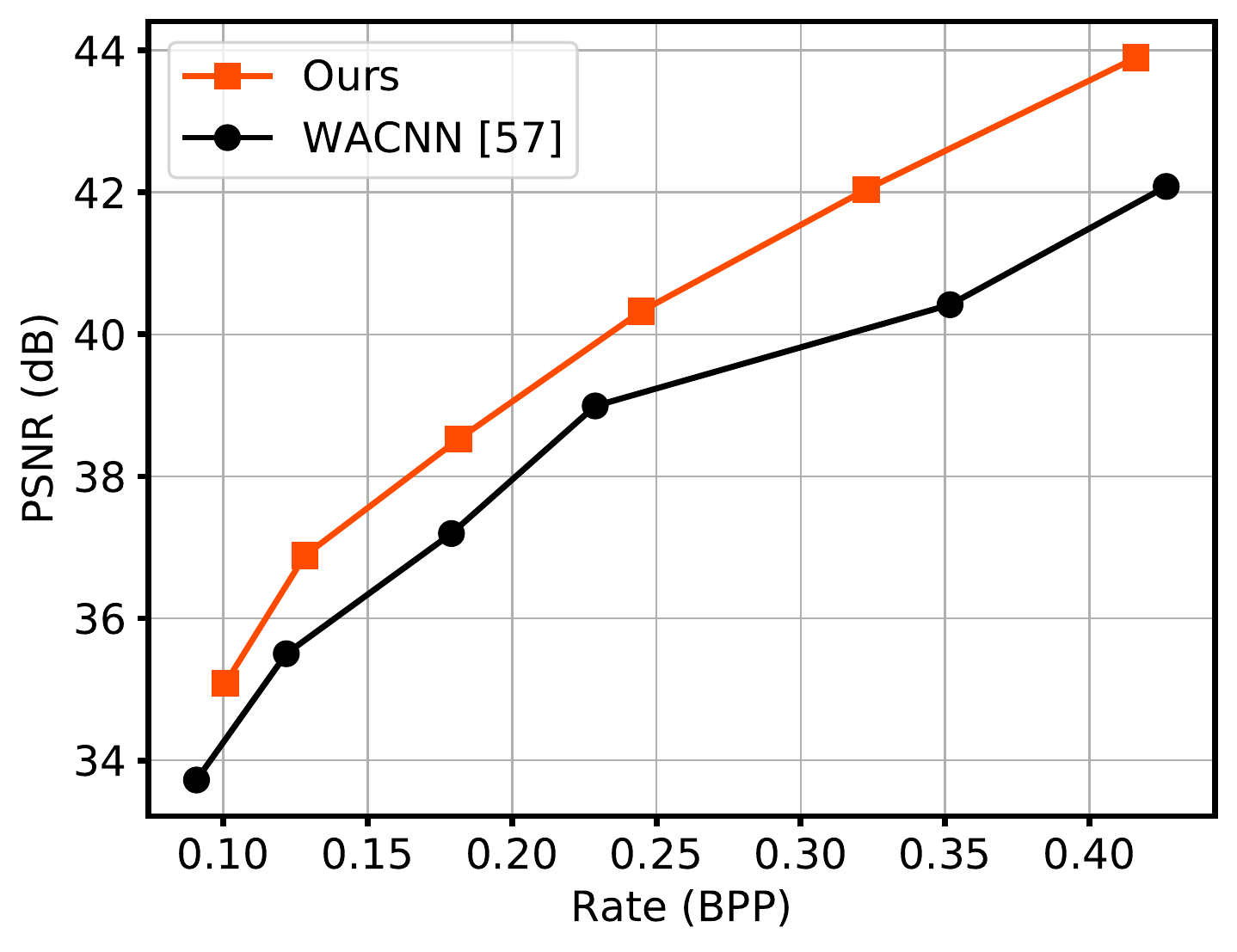}
    \vskip-5pt
    \subcaption*{Vector art}
  \end{minipage}
  \caption{Comparison with WACNN~\cite{ZouCVPR22}, which is the baseline method that does not perform adaptive optimization.}
  \label{fig:results}
  \vspace{-8pt}
\end{figure*}

\begin{table*}[t]
  \centering
  \caption{Comparison with existing adaptive compression methods on BD rate (\%) to VVC~\cite{VVC}. The BD rates of JPEG, BPG, VVC, and WACNN~\cite{ZouCVPR22} are provided for reference. A smaller value is more effective.}
  \label{tbl:result}
  \begin{tabular}{lccccc}
    \toprule
    Method & Natural image & Comic & Line drawing & Vector art & Average\\
    \midrule
    JPEG & 184 &       447 &         186 &           676 &   373\\
    BPG &  33.7 &        88.0 &          28.8 &           114 &    66.2 \\
    VVC & 0.00 & 0.00 & 0.00 & 0.00 & 0.00\\
    WACNN~\cite{ZouCVPR22} &  -6.31 &        11.6 &          14.5 &            25.3 &    11.3 \\
    \midrule
    Yang \etal~\cite{conf/nips/Yang20} &  \textbf{-9.82} &        -0.50 &           1.84 &             8.47 &    -0.00\\
    Lam \etal~\cite{conf/mm/LamZCLH20} &  151 &        197 &          161 &            367 &    219\\
    Rozendaal \etal~\cite{rozendaal2021overfitting} & 234 & 317 & 267 & 718 & 384\\
    Zou \etal~\cite{ZouAdaptISM21} &  -9.68 &        -2.40 &          -0.13 &             4.12 &    -2.02 \\
    Ours            &          -9.79 &        \textbf{-2.82} &          \textbf{-0.25} &             \textbf{2.87} &    \textbf{-2.50}\\
    \bottomrule
  \end{tabular}
  \vspace{-8pt}
\end{table*}

\begin{figure*}[t]
  \begin{minipage}{0.19\hsize}
    \centering
    \centering
    \includegraphics[width=\hsize]{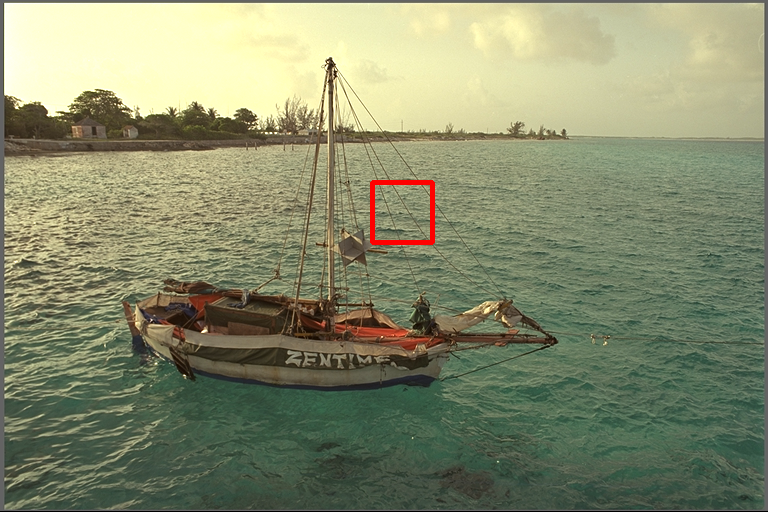}
    \captionsetup{justification=centering}
    \subcaption*{Natural Image}
  \end{minipage}
  \begin{minipage}{0.13\hsize}
    \centering
    \setlength{\fboxrule}{1pt}
    \setlength{\fboxsep}{0pt}
    \fcolorbox{red}{white}{\includegraphics[width=\hsize]{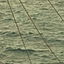}}
    \captionsetup{justification=centering}
    \subcaption*{PSNR/BPP}
  \end{minipage}
  \begin{minipage}{0.13\hsize}
    \centering
    \setlength{\fboxrule}{1pt}
    \setlength{\fboxsep}{0pt}
    \fcolorbox{red}{white}{\includegraphics[width=\hsize]{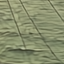}}
    \captionsetup{justification=centering}
    \subcaption*{\textbf{29.6}/0.242}
  \end{minipage}
  \begin{minipage}{0.13\hsize}
    \centering
    \setlength{\fboxrule}{1pt}
    \setlength{\fboxsep}{0pt}
    \fcolorbox{red}{white}{\includegraphics[width=\hsize]{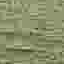}}
    \captionsetup{justification=centering}
    \subcaption*{23.7/0.252}
  \end{minipage}
  \begin{minipage}{0.13\hsize}
    \centering
    \setlength{\fboxrule}{1pt}
    \setlength{\fboxsep}{0pt}
    \fcolorbox{red}{white}{\includegraphics[width=\hsize]{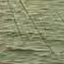}}
    \captionsetup{justification=centering}
    \subcaption*{28.7/0.254}
  \end{minipage}
  \begin{minipage}{0.13\hsize}
    \centering
    \setlength{\fboxrule}{1pt}
    \setlength{\fboxsep}{0pt}
    \fcolorbox{red}{white}{\includegraphics[width=\hsize]{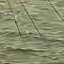}}
    \captionsetup{justification=centering}
    \subcaption*{29.5/0.254}
  \end{minipage}
  \begin{minipage}{0.13\hsize}
    \centering
    \setlength{\fboxrule}{1pt}
    \setlength{\fboxsep}{0pt}
    \fcolorbox{red}{white}{\includegraphics[width=\hsize]{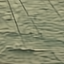}}
    \captionsetup{justification=centering}
    \subcaption*{29.3/0.247}
  \end{minipage}

  \begin{minipage}{0.19\hsize}
    \centering
    \centering
    \includegraphics[width=0.8\hsize]{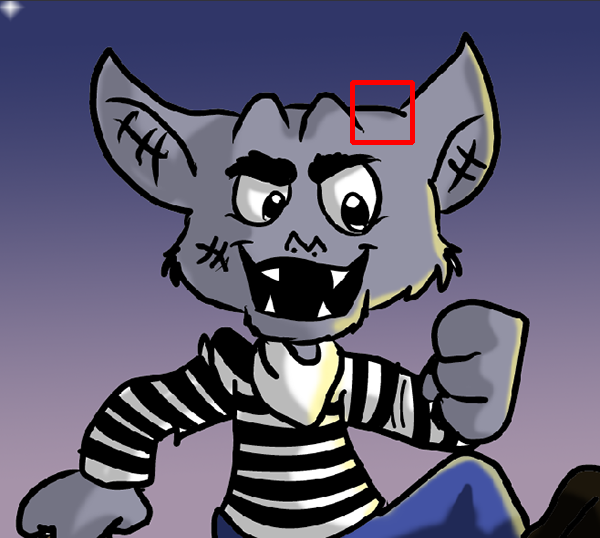}
    \captionsetup{justification=centering}
    \subcaption*{Comic}
  \end{minipage}
  \begin{minipage}{0.13\hsize}
    \centering
    \setlength{\fboxrule}{1pt}
    \setlength{\fboxsep}{0pt}
    \fcolorbox{red}{white}{\includegraphics[width=\hsize]{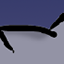}}
    \captionsetup{justification=centering}
    \subcaption*{PSNR/BPP}
  \end{minipage}
  \begin{minipage}{0.13\hsize}
    \centering
    \setlength{\fboxrule}{1pt}
    \setlength{\fboxsep}{0pt}
    \fcolorbox{red}{white}{\includegraphics[width=\hsize]{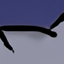}}
    \captionsetup{justification=centering}
    \subcaption*{\textbf{37.4}/0.171}
  \end{minipage}
  \begin{minipage}{0.13\hsize}
    \centering
    \setlength{\fboxrule}{1pt}
    \setlength{\fboxsep}{0pt}
    \fcolorbox{red}{white}{\includegraphics[width=\hsize]{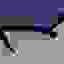}}
    \captionsetup{justification=centering}
    \subcaption*{21.9/0.212}
  \end{minipage}
  \begin{minipage}{0.13\hsize}
    \centering
    \setlength{\fboxrule}{1pt}
    \setlength{\fboxsep}{0pt}
    \fcolorbox{red}{white}{\includegraphics[width=\hsize]{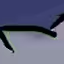}}
    \captionsetup{justification=centering}
    \subcaption*{32.6/0.177}
  \end{minipage}
  \begin{minipage}{0.13\hsize}
    \centering
    \setlength{\fboxrule}{1pt}
    \setlength{\fboxsep}{0pt}
    \fcolorbox{red}{white}{\includegraphics[width=\hsize]{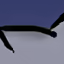}}
    \captionsetup{justification=centering}
    \subcaption*{37.2/0.183}
  \end{minipage}
  \begin{minipage}{0.13\hsize}
    \centering
    \setlength{\fboxrule}{1pt}
    \setlength{\fboxsep}{0pt}
    \fcolorbox{red}{white}{\includegraphics[width=\hsize]{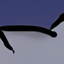}}
    \captionsetup{justification=centering}
    \subcaption*{36.7/0.180}
  \end{minipage}

  \begin{minipage}{0.19\hsize}
    \centering
    \centering
    \includegraphics[width=\hsize]{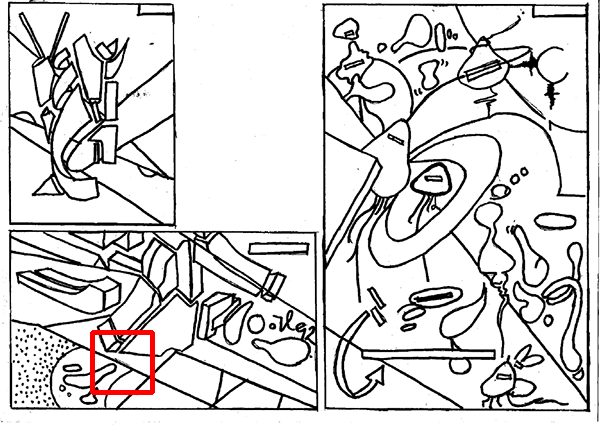}
    \captionsetup{justification=centering}
    \subcaption*{Line drawing}
  \end{minipage}
  \begin{minipage}{0.13\hsize}
    \centering
    \setlength{\fboxrule}{1pt}
    \setlength{\fboxsep}{0pt}
    \fcolorbox{red}{white}{\includegraphics[width=\hsize]{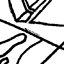}}
    \captionsetup{justification=centering}
    \subcaption*{PSNR/BPP}
  \end{minipage}
  \begin{minipage}{0.13\hsize}
    \centering
    \setlength{\fboxrule}{1pt}
    \setlength{\fboxsep}{0pt}
    \fcolorbox{red}{white}{\includegraphics[width=\hsize]{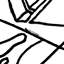}}
    \captionsetup{justification=centering}
    \subcaption*{\textbf{30.9}/0.396}
  \end{minipage}
  \begin{minipage}{0.13\hsize}
    \centering
    \setlength{\fboxrule}{1pt}
    \setlength{\fboxsep}{0pt}
    \fcolorbox{red}{white}{\includegraphics[width=\hsize]{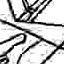}}
    \captionsetup{justification=centering}
    \subcaption*{19.7/0.459}
  \end{minipage}
  \begin{minipage}{0.13\hsize}
    \centering
    \setlength{\fboxrule}{1pt}
    \setlength{\fboxsep}{0pt}
    \fcolorbox{red}{white}{\includegraphics[width=\hsize]{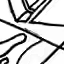}}
    \captionsetup{justification=centering}
    \subcaption*{27.2/0.420}
  \end{minipage}
  \begin{minipage}{0.13\hsize}
    \centering
    \setlength{\fboxrule}{1pt}
    \setlength{\fboxsep}{0pt}
    \fcolorbox{red}{white}{\includegraphics[width=\hsize]{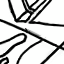}}
    \captionsetup{justification=centering}
    \subcaption*{30.2/0.387}
  \end{minipage}
  \begin{minipage}{0.13\hsize}
    \centering
    \setlength{\fboxrule}{1pt}
    \setlength{\fboxsep}{0pt}
    \fcolorbox{red}{white}{\includegraphics[width=\hsize]{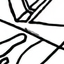}}
    \captionsetup{justification=centering}
    \subcaption*{29.4/0.391}
  \end{minipage}

  \begin{minipage}{0.19\hsize}
    \centering
    \centering
    \includegraphics[width=0.6\hsize]{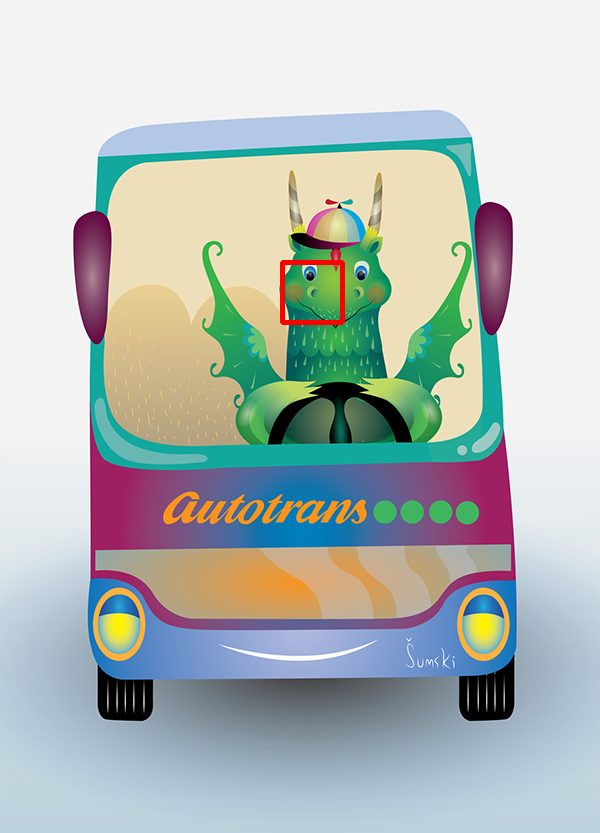}
    \captionsetup{justification=centering}
    \subcaption*{Vector art}
  \end{minipage}
  \begin{minipage}{0.13\hsize}
    \centering
    \setlength{\fboxrule}{1pt}
    \setlength{\fboxsep}{0pt}
    \fcolorbox{red}{white}{\includegraphics[width=\hsize]{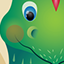}}
    \captionsetup{justification=centering}
    \subcaption*{PSNR/BPP\\Input}
  \end{minipage}
  \begin{minipage}{0.13\hsize}
    \centering
    \setlength{\fboxrule}{1pt}
    \setlength{\fboxsep}{0pt}
    \fcolorbox{red}{white}{\includegraphics[width=\hsize]{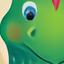}}
    \captionsetup{justification=centering}
    \subcaption*{\textbf{33.7}/0.071\\Ours}
  \end{minipage}
  \begin{minipage}{0.13\hsize}
    \centering
    \setlength{\fboxrule}{1pt}
    \setlength{\fboxsep}{0pt}
    \fcolorbox{red}{white}{\includegraphics[width=\hsize]{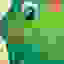}}
    \captionsetup{justification=centering}
    \subcaption*{21.3/0.169\\JPEG}
  \end{minipage}
  \begin{minipage}{0.13\hsize}
    \centering
    \setlength{\fboxrule}{1pt}
    \setlength{\fboxsep}{0pt}
    \fcolorbox{red}{white}{\includegraphics[width=\hsize]{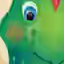}}
    \captionsetup{justification=centering}
    \subcaption*{29.4/0.074\\BPG}
  \end{minipage}
  \begin{minipage}{0.13\hsize}
    \centering
    \setlength{\fboxrule}{1pt}
    \setlength{\fboxsep}{0pt}
    \fcolorbox{red}{white}{\includegraphics[width=\hsize]{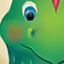}}
    \captionsetup{justification=centering}
    \subcaption*{33.3/0.069\\VVC}
  \end{minipage}
  \begin{minipage}{0.13\hsize}
    \centering
    \setlength{\fboxrule}{1pt}
    \setlength{\fboxsep}{0pt}
    \fcolorbox{red}{white}{\includegraphics[width=\hsize]{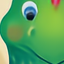}}
    \captionsetup{justification=centering}
    \subcaption*{32.4/0.064\\WACNN~\cite{ZouCVPR22}}
  \end{minipage}
  \caption{Qualitative results for the four domains. Ours reconstruct the wires, shadow of the character's head, overlapping lines, and the monster's nose with relatively high fidelity, respectively.}
  \label{fig:qualitative}
  \vspace{-8pt}
\end{figure*}

\subsection{Comparison with Other Methods}
\textbf{Rate-Distortion Performance.}
First, we compared the proposed method with a baseline method that does not perform adaptive optimization.
We calculated the peak signal-to-noise ratio (PSNR) and bits per pixel (BPP) for each image, and computed the average values to plot on a rate-distortion curve.
The results are displayed in Fig.~\ref{fig:results}.
Evidently, the proposed method significantly outperformed the baseline method.
The improvement in PSNR was approximately 1--2 dB.
This indicates that adaptive optimization is effective for universal deep image compression.

Next, we compared the proposed method with other adaptation methods.
For a fair comparison, we reimplemented Yang \etal~\cite{conf/nips/Yang20}, Rozendaal \etal~\cite{rozendaal2021overfitting}, Zou \etal~\cite{ZouAdaptISM21}, and Lam \etal~\cite{conf/mm/LamZCLH20} in our framework.
Please refer to the supplementary material for the detailed experimental setup.
Additionally, we performed a comparison with the baseline method and three conventional codecs: JPEG~\cite{JPEG}, BPG~\cite{BPG}, and VVC~\cite{VVC} for reference.
In particular, we used VVC for intra-frames implemented in VTM~\cite{VTM}.
We computed Bj{\o}ntegaard Delta bitrate (BD rate)~\cite{BDrate} compared with VVC.
The results are presented in Table~\ref{tbl:result}.
Evidently, the proposed method achieved performance superior to those of the other adaptation methods.
Furthermore, the proposed method outperformed VVC, which is the state-of-the-art conventional method.
Note that Lam \etal~\cite{conf/mm/LamZCLH20} and Rozendaal \etal~\cite{rozendaal2021overfitting} performed inferior to the baseline method.
This is because these methods transmitted many parameters in the decoder for an individual image.

\textbf{Qualitative Results.}
The qualitative results of the proposed method, baseline method, and conventional codecs are shown in Fig.~\ref{fig:qualitative}.
We compared these methods at a similar BPP.
Our method achieved higher visual quality compared with conventional codecs and the baseline method.

\textbf{Runtime.}
We measured the runtime of encoding and decoding using GPUs (NVIDIA Tesla V100).
We conducted experiments on vector arts using the baseline and proposed methods.
We show the average runtime in Table~\ref{tbl:runtime}.
The decoding time was found to be comparable to that of the baseline method.
However, the proposed method required more time for encoding than the baseline method owing to the adaptive optimization framework.

\begin{table}[t]
  \centering
  \caption{Comparison of runtime.}
  \label{tbl:runtime}
  \begin{tabular}{lcc}
    \toprule
    Method & Encoding (s) & Decoding (s)\\
    \midrule
    WACNN~\cite{ZouCVPR22} & 0.16 & 0.16\\
    Ours & 260 & 0.18\\
    \bottomrule
  \end{tabular}
  \vspace{-8pt}
\end{table}

\subsection{Ablation Studies}
\textbf{Effectiveness of Adapters.}
To show the effectiveness of adapters, we compared the proposed method with other methods that update parameters other than adapter parameters.
In the experiments, we updated zero parameters, biases of the layers as in \cite{conf/mm/LamZCLH20}, and overfittable multiplicative parameters (OMPs) as in \cite{ZouAdaptISM21}.
The numbers of updated parameters were 0, 9283, and 192, respectively.

\begin{figure}[t]
  \centering
  \begin{minipage}{\hsize}
    \centering
    \includegraphics[width=0.8\hsize]{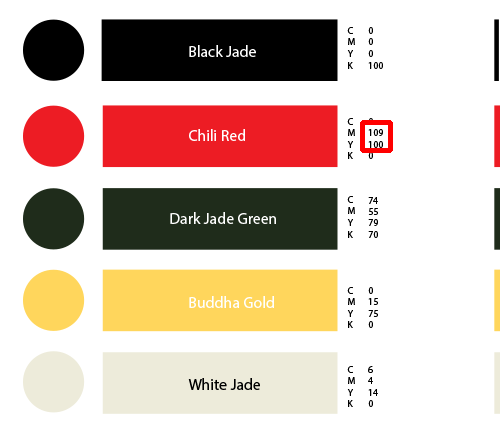}
    \captionsetup{justification=centering}
  \end{minipage}\\
  
  \begin{minipage}{0.32\hsize}
    \centering
    \setlength{\fboxrule}{1pt}
    \setlength{\fboxsep}{0pt}
    \fcolorbox{red}{white}{\includegraphics[width=\hsize]{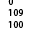}}
    \captionsetup{justification=centering}
    \subcaption*{PSNR/BPP\\Input}
  \end{minipage}
  \begin{minipage}{0.32\hsize}
    \centering
    \setlength{\fboxrule}{1pt}
    \setlength{\fboxsep}{0pt}
    \fcolorbox{red}{white}{\includegraphics[width=\hsize]{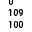}}
    \captionsetup{justification=centering}
    \subcaption*{35.4/0.099\\Ours}
  \end{minipage}
  \begin{minipage}{0.32\hsize}
    \centering
    \setlength{\fboxrule}{1pt}
    \setlength{\fboxsep}{0pt}
    \fcolorbox{red}{white}{\includegraphics[width=\hsize]{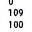}}
    \captionsetup{justification=centering}
    \subcaption*{34.5/0.090\\Without adapters}
  \end{minipage}
  \caption{Qualitative results for the effectiveness of adapters. The top image is the entire input image, whereas the bottom images are the cropped patches. We can observe artifacts around the texts are reduced by using adapters.}
  \label{fig:abl_adpt}
  \vspace{-8pt}
\end{figure}

The results on vector arts are listed in Table~\ref{tbl:abl_adpt} and revealed that the highest performance is obtained when adapters are adapted.
The qualitative results are shown in Fig.~\ref{fig:abl_adpt}.
Evidently, the artifacts around the boundary of the texts were reduced using adapters.

\begin{table}[t]
  \centering
  \caption{Comparison of update of different parameters on BD rate (\%). A smaller value is more effective.}
  \label{tbl:abl_adpt}
  \begin{tabular}{lcc}
    \toprule
    Method & \# of parameters & BD rate (\%) $\downarrow$\\
    \midrule
    Adapters (Ours) & 768 & \textbf{0.00}\\
    Zero parameters & 0 & 6.16\\
    Biases & 9283 & 42.1\\
    OMPs & 192 & 0.91\\
    \bottomrule
  \end{tabular}
  \vspace{-8pt}
\end{table}

\textbf{Effectiveness of Rate-Distortion Optimization of Adapters.}
In our framework, we optimized the adapters in terms of rate-distortion.
In this experiment, we present the results for when the adapters were optimized only in terms of distortion and compressed into eight bits as in Zou \etal~\cite{ZouAdaptISM21}.
In the implementation, we linearly transformed the parameters of the adapters to the range of $[0, 255]$ and quantized the parameters to the integer.
This was performed to obtain integer values of eight bits and two real values of 32 bits, which are the scale and bias for the linear transformation.
The BD rate compared with our method on vector arts was 4.21\%.
The results indicate the effectiveness of the rate-distortion optimization of the adapters.

\begin{figure}[t]
  \centering
  \includegraphics[width=0.95\hsize]{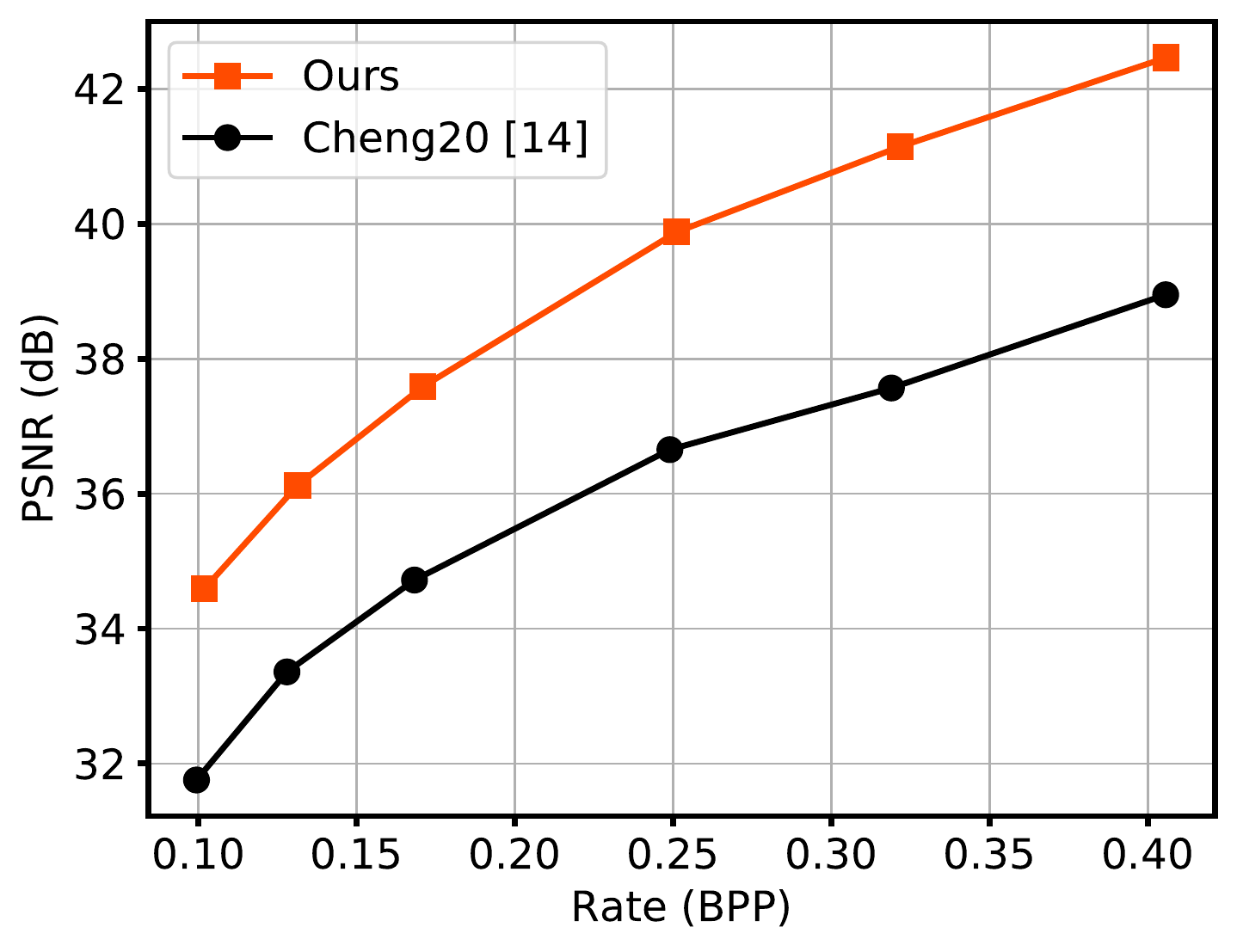}
  \vskip-7pt
  \caption{Results on vector arts using Cheng20~\cite{conf/cvpr/Cheng2020} as base network architecture.}
  \label{fig:cheng20}
  \vspace{-8pt}
\end{figure}

\textbf{Application to Another Network Architecture.}
Our framework can be applied to other network architectures.
In this experiment, we demonstrate the performance of our framework when it is applied to Cheng20~\cite{conf/cvpr/Cheng2020}.
In our implementation, we used cheng2020-attn in CompressAI~\cite{compressai}.
We used publicly available models pre-trained on natural images.
Subsequently, we inserted adapters after the convolutional layer at the first side of the final residual block of cheng2020-attn.
The results are shown in Fig.~\ref{fig:cheng20}, which revealed that our framework significantly outperformed the baseline method on Cheng20~\cite{conf/cvpr/Cheng2020}.

\textbf{Optimization Order.}
Our framework first optimizes the latent representation and then optimizes the parameters of the adapters.
In this experiment, we swap the order of the optimization.
That is, we first train the adapters and then refine the latent representation using the trained adapters.
The BD rate compared with our method on vector arts was 0.70\%.
The results indicate that our optimization order is effective.

\section{Conclusion}
In this study, we addressed a novel task that we named universal deep image compression.
We observed a problem wherein deep image compression deteriorates its performance on out-of-domain images.
We proposed a content-adaptive optimization framework to address this problem.
To adapt a pre-trained compression model per target image, we refined the latent representation extracted by the encoder and trained the adapters inserted into the decoder.
Our framework can be applied to all pre-trained compression models.
We constructed a benchmark dataset with four domains and demonstrated that our framework is effective.
The limitation of our research is an expensive encoding time due to the optimization during compression.
Reducing the encoding time is an important future work of our research.

\textbf{Acknowledgement.}
This work was partially supported by JSPS KAKENHI Grant Number 22J13735 and 21H03460.

\clearpage
{\small
\bibliographystyle{ieee_fullname}
\bibliography{egbib}
}

\clearpage

\appendix
\onecolumn
\section{Supplementary Material}
This supplementary material provides (1) a detailed experimental setup of existing adaptive compression methods and (2) whole images of the qualitative results that we are not able to show in the main paper.

\subsection{Detailed Experimental Setup of Existing Adaptive Compression Methods}
We describe the detailed experimental setup of four existing adaptive compression methods: (1) Yang \etal~\cite{conf/nips/Yang20}, (2) Lam \etal~\cite{conf/mm/LamZCLH20}, (3) Rozendaal \etal~\cite{rozendaal2021overfitting}, and (4) Zou \etal~\cite{ZouAdaptISM21}.
(1) Yang \etal performs latent refinement without any adapter, which is equivalent to the first stage of our approach.
(2) Lam \etal updates the parameters of the biases of the convolutional layers in the decoder after the latent refinement.
The parameters are optimized only in terms of distortion.
The updated parameters are converted to 64 bits floating point numbers and compressed in the 7z format.
The number of updated parameters is 9283.
(3) Rozendaal \etal updates all the parameters in the decoder and entropy model.
The parameters are optimized in terms of rate-distortion.
The number of updated parameters is $6.50 \times 10^7$.
(4) Lam \etal uses overfittable multiplicative parameters (OMPs) instead of adapters after the latent refinement.
The parameters are optimized only in terms of distortion.
The updated parameters are transformed linearly to the range of [0, 255] and quantized to the integer.
This was performed to obtain integer values of eight bits and two real values of 32 bits, which are the scale and bias for the linear transformation.
The number of updated parameters is 192.

\subsection{Whole Images of Qualitative Results}
\begin{figure*}[b]
  \begin{minipage}{0.32\hsize}
    \centering
    \includegraphics[width=\hsize]{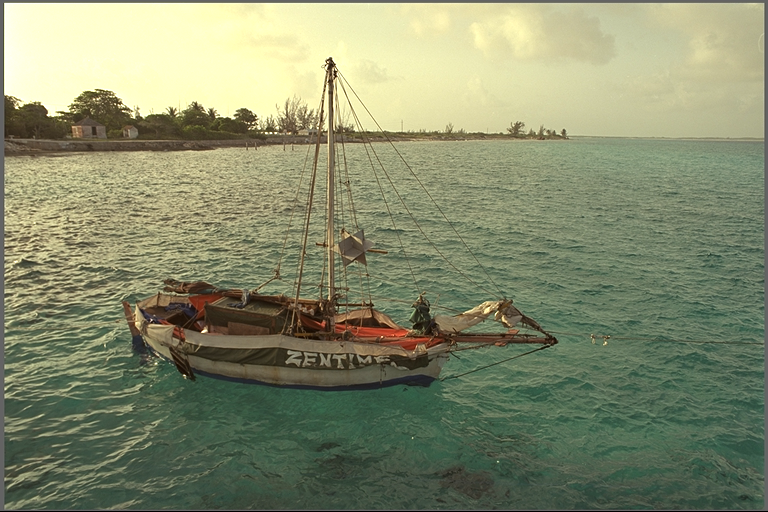}
    \captionsetup{justification=centering}
    \subcaption*{Input (Natual image)\\PSNR/BPP}
  \end{minipage}
  \begin{minipage}{0.32\hsize}
    \centering
    \includegraphics[width=\hsize]{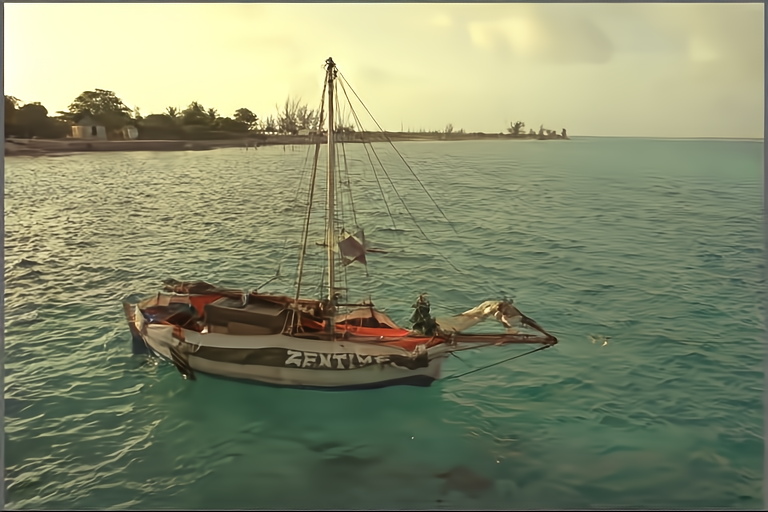}
    \captionsetup{justification=centering}
    \subcaption*{Ours\\\textbf{29.6}/0.242}
  \end{minipage}
  \begin{minipage}{0.32\hsize}
    \centering
    \includegraphics[width=\hsize]{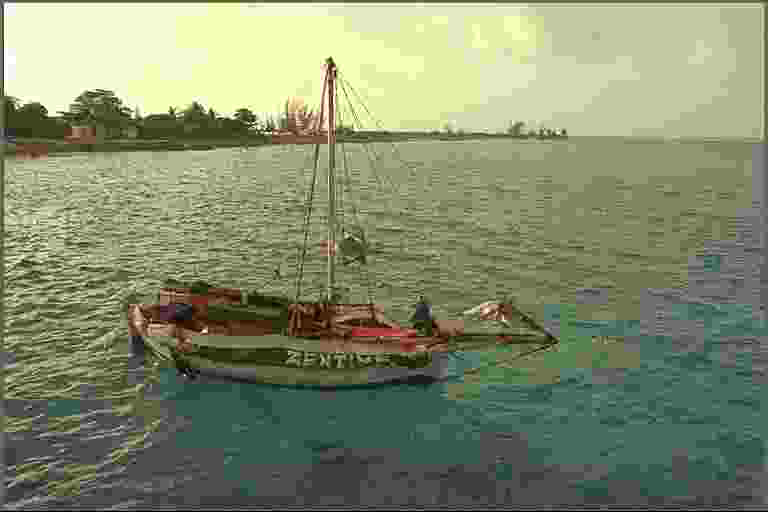}
    \captionsetup{justification=centering}
    \subcaption*{JPEG\\23.7/0.252}
  \end{minipage}

  \begin{minipage}{0.32\hsize}
    \centering
    \includegraphics[width=\hsize]{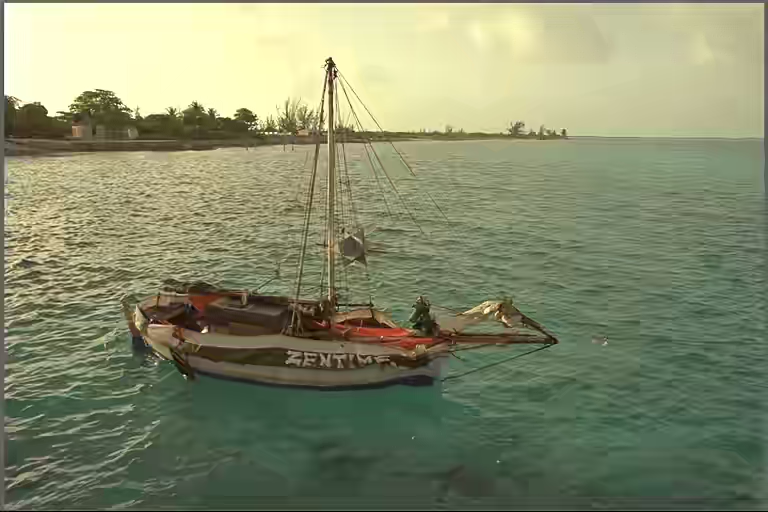}
    \captionsetup{justification=centering}
    \subcaption*{BPG\\28.7/0.254}
  \end{minipage}
  \begin{minipage}{0.32\hsize}
    \centering
    \includegraphics[width=\hsize]{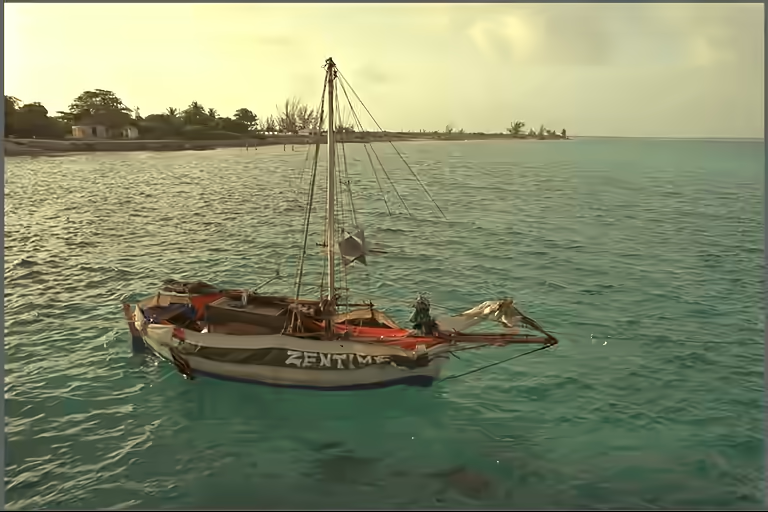}
    \captionsetup{justification=centering}
    \subcaption*{VVC\\29.5/0.254}
  \end{minipage}
  \begin{minipage}{0.32\hsize}
    \centering
    \includegraphics[width=\hsize]{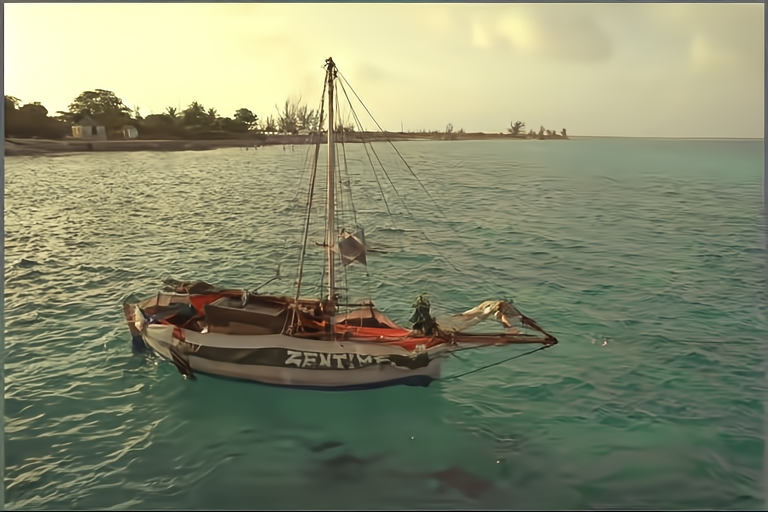}
    \captionsetup{justification=centering}
    \subcaption*{WACNN~\cite{ZouCVPR22}\\29.3/0.247}
  \end{minipage}
  \caption{Qualitative comparison with other compression methods on a natural image.}
  \label{fig:qua_natural}
\end{figure*}

\begin{figure*}[t]
  \begin{minipage}{0.32\hsize}
    \centering
    \includegraphics[width=\hsize]{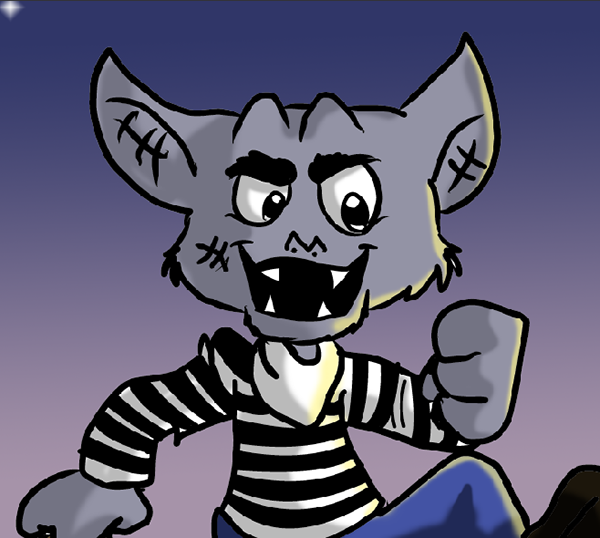}
    \captionsetup{justification=centering}
    \subcaption*{Input (Comic)\\PSNR/BPP}
  \end{minipage}
  \begin{minipage}{0.32\hsize}
    \centering
    \includegraphics[width=\hsize]{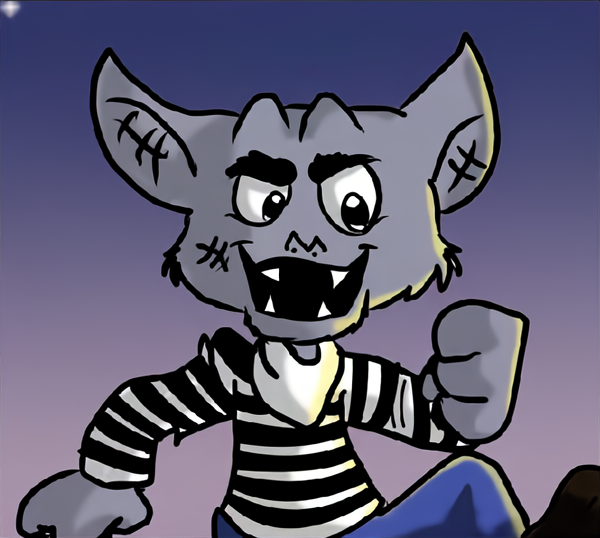}
    \captionsetup{justification=centering}
    \subcaption*{Ours\\\textbf{37.4}/0.171}
  \end{minipage}
  \begin{minipage}{0.32\hsize}
    \centering
    \includegraphics[width=\hsize]{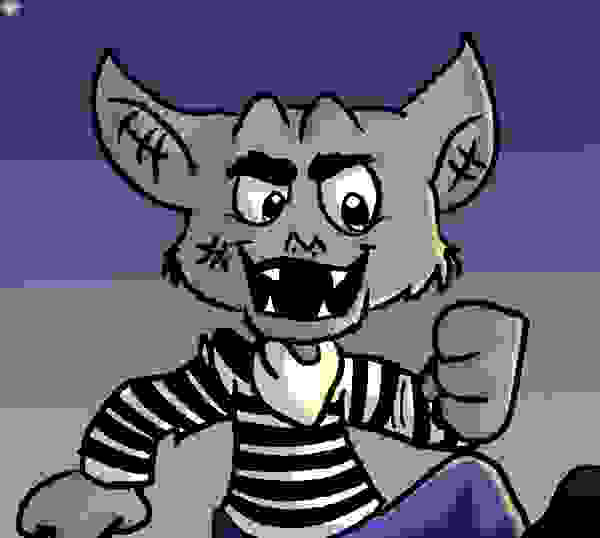}
    \captionsetup{justification=centering}
    \subcaption*{JPEG\\21.9/0.212}
  \end{minipage}
  
  \begin{minipage}{0.32\hsize}
    \centering
    \includegraphics[width=\hsize]{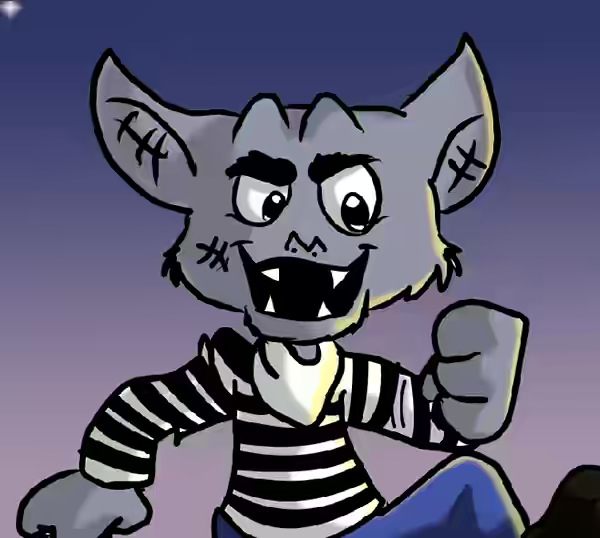}
    \captionsetup{justification=centering}
    \subcaption*{BPG\\32.6/0.177}
  \end{minipage}
  \begin{minipage}{0.32\hsize}
    \centering
    \includegraphics[width=\hsize]{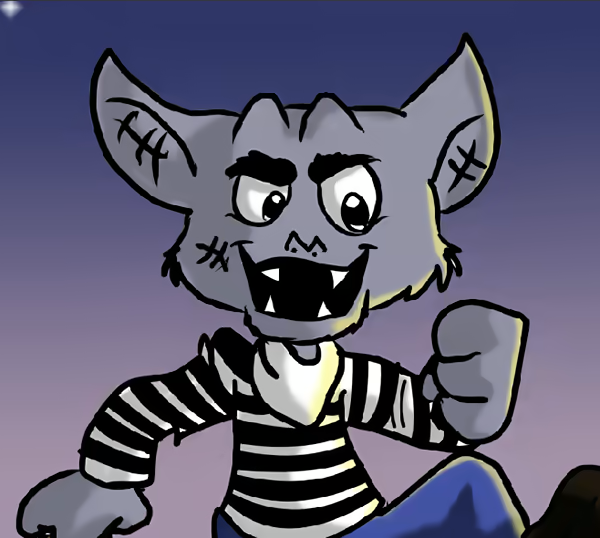}
    \captionsetup{justification=centering}
    \subcaption*{VVC\\37.2/0.183}
  \end{minipage}
  \begin{minipage}{0.32\hsize}
    \centering
    \includegraphics[width=\hsize]{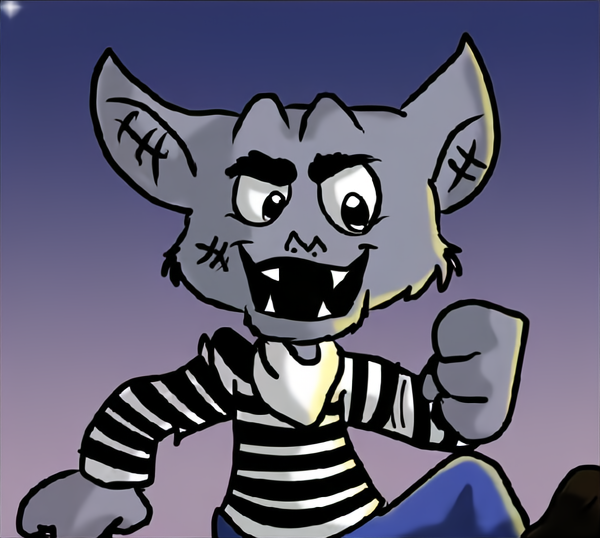}
    \captionsetup{justification=centering}
    \subcaption*{WACNN~\cite{ZouCVPR22}\\36.7/0.180}
  \end{minipage}
  \caption{Qualitative comparison with other compression methods on a comic image.}
  \label{fig:qua_comic}
\end{figure*}

\begin{figure*}[t]
  \begin{minipage}{0.32\hsize}
    \centering
    \includegraphics[width=\hsize]{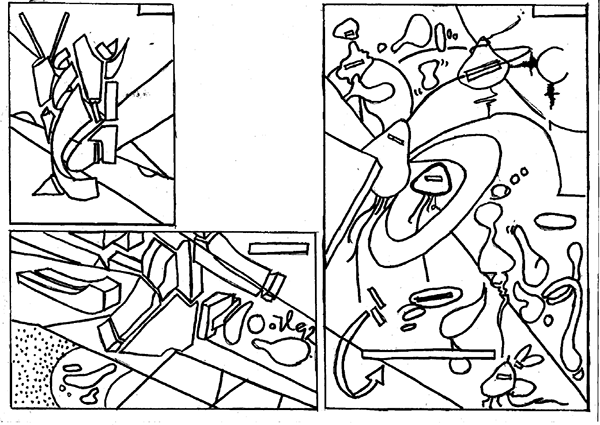}
    \captionsetup{justification=centering}
    \subcaption*{Input (Line drawing)\\PSNR/BPP}
  \end{minipage}
  \begin{minipage}{0.32\hsize}
    \centering
    \includegraphics[width=\hsize]{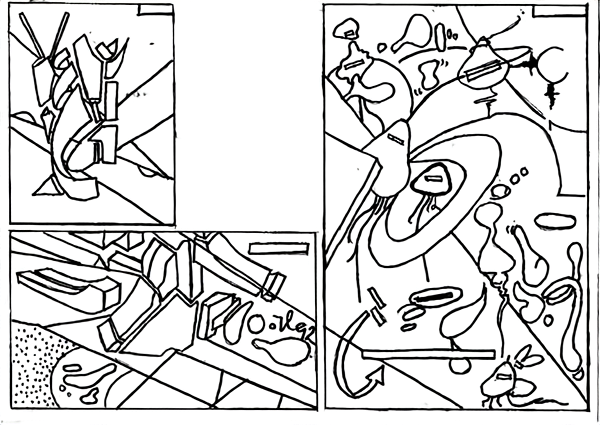}
    \captionsetup{justification=centering}
    \subcaption*{Ours\\\textbf{30.9}/0.396}
  \end{minipage}
  \begin{minipage}{0.32\hsize}
    \centering
    \includegraphics[width=\hsize]{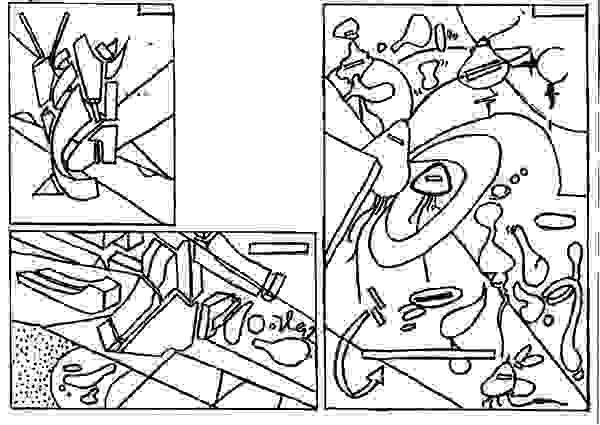}
    \captionsetup{justification=centering}
    \subcaption*{JPEG\\19.7/0.459}
  \end{minipage}
  
  \begin{minipage}{0.32\hsize}
    \centering
    \includegraphics[width=\hsize]{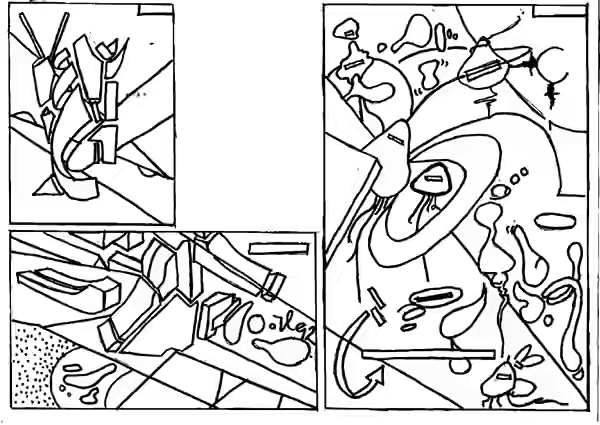}
    \captionsetup{justification=centering}
    \subcaption*{BPG\\27.2/0.420}
  \end{minipage}
  \begin{minipage}{0.32\hsize}
    \centering
    \includegraphics[width=\hsize]{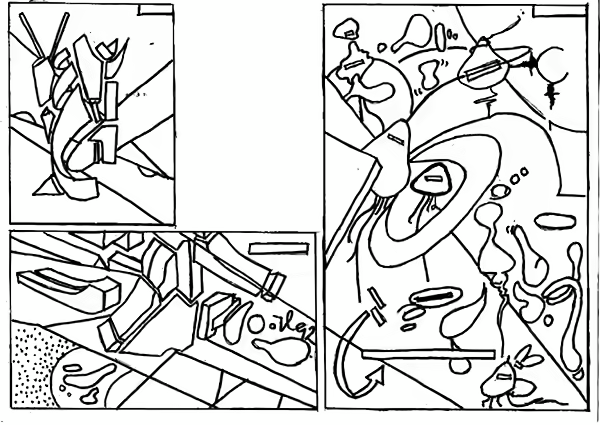}
    \captionsetup{justification=centering}
    \subcaption*{VVC\\30.2/0.387}
  \end{minipage}
  \begin{minipage}{0.32\hsize}
    \centering
    \includegraphics[width=\hsize]{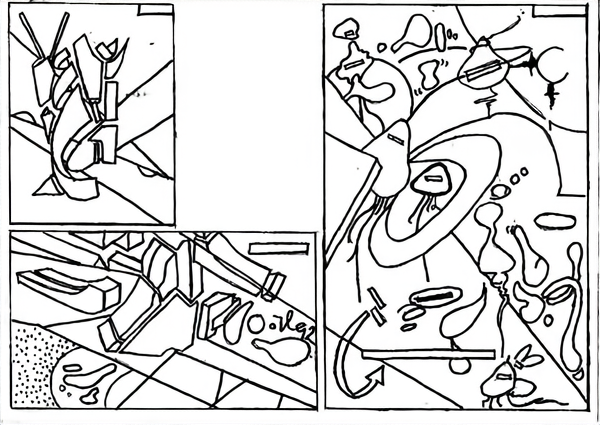}
    \captionsetup{justification=centering}
    \subcaption*{WACNN~\cite{ZouCVPR22}\\29.4/0.391}
  \end{minipage}
  \caption{Qualitative comparison with other compression methods on a line drawing.}
  \label{fig:qua_pen}
\end{figure*}

\begin{figure*}[t]
  \begin{minipage}{0.32\hsize}
    \centering
    \centering
    \includegraphics[width=\hsize]{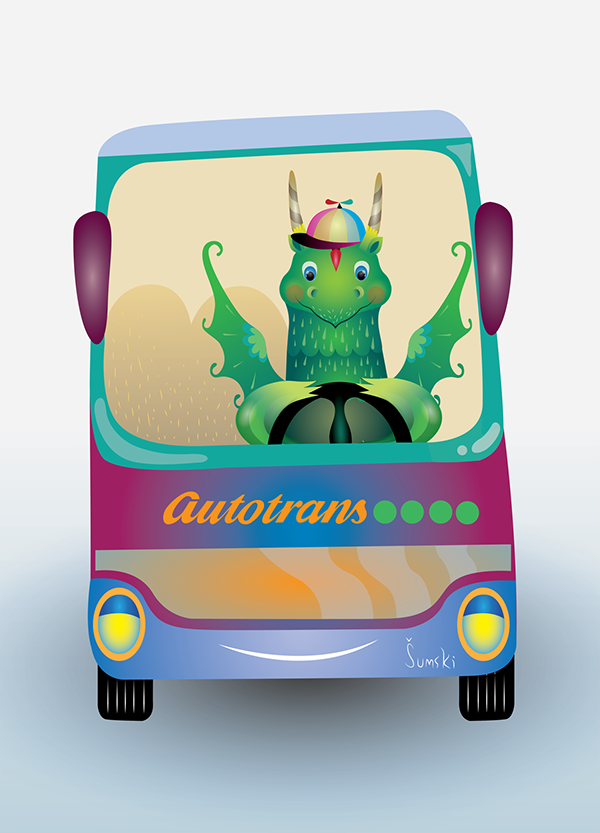}
    \captionsetup{justification=centering}
    \subcaption*{Input (Vector art)\\PSNR/BPP}
  \end{minipage}
  \begin{minipage}{0.32\hsize}
    \centering
    \includegraphics[width=\hsize]{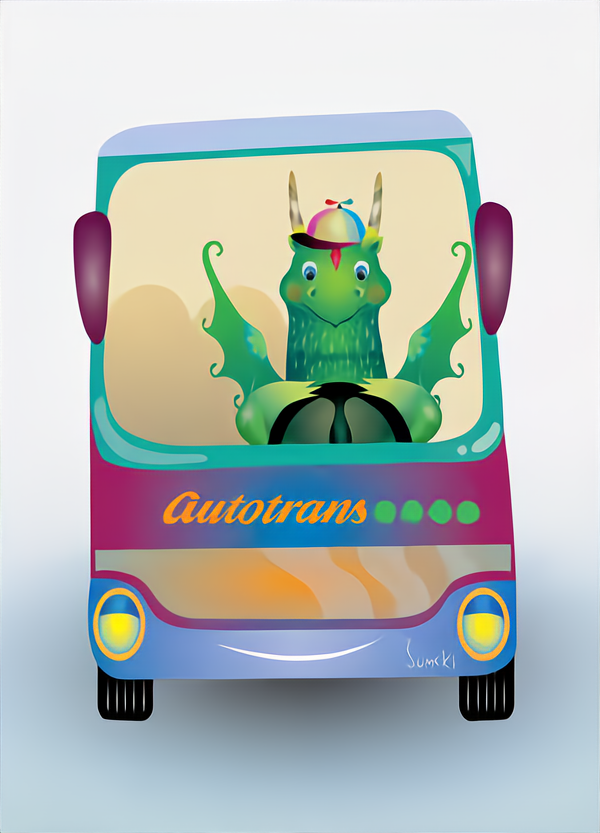}
    \captionsetup{justification=centering}
    \subcaption*{Ours\\\textbf{33.7}/0.071}
  \end{minipage}
  \begin{minipage}{0.32\hsize}
    \centering
    \includegraphics[width=\hsize]{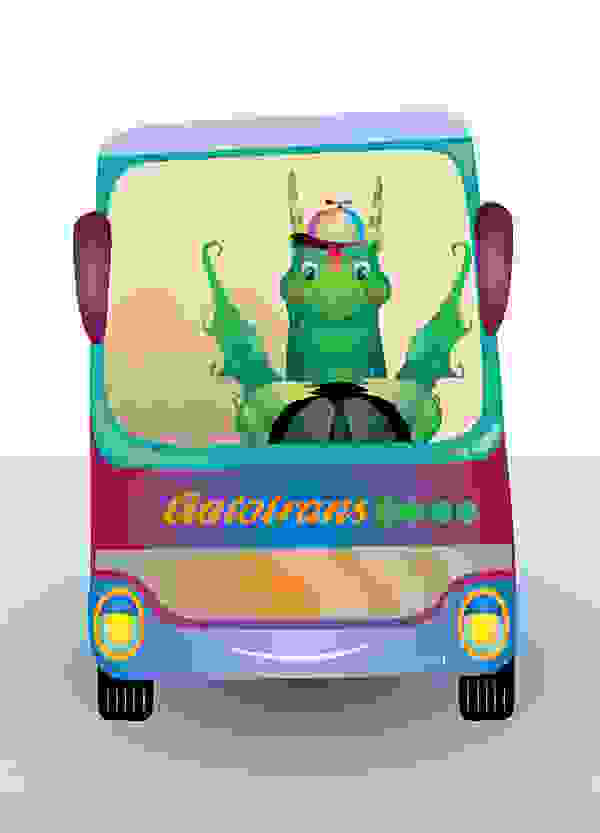}
    \captionsetup{justification=centering}
    \subcaption*{JPEG\\21.3/0.169}
  \end{minipage}

  \begin{minipage}{0.32\hsize}
    \centering
    \includegraphics[width=\hsize]{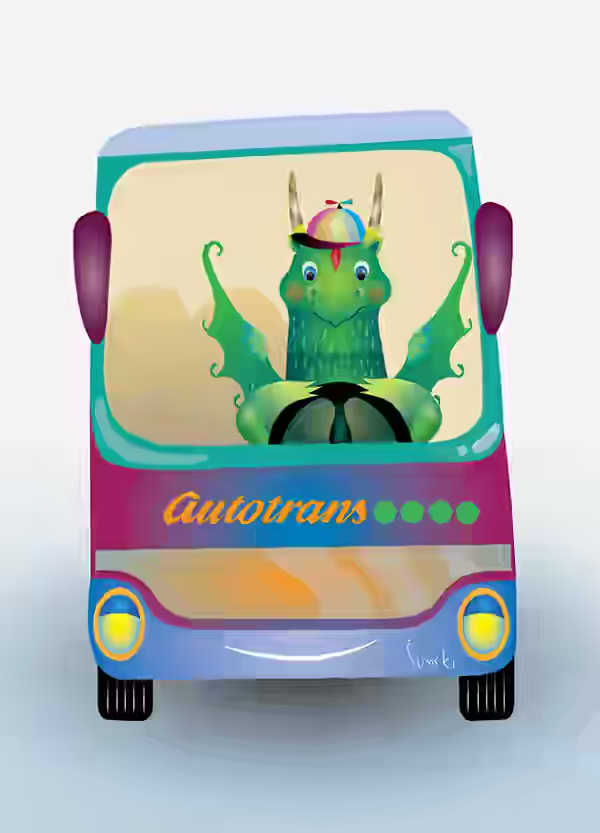}
    \captionsetup{justification=centering}
    \subcaption*{BPG\\29.4/0.074}
  \end{minipage}
  \begin{minipage}{0.32\hsize}
    \centering
    \includegraphics[width=\hsize]{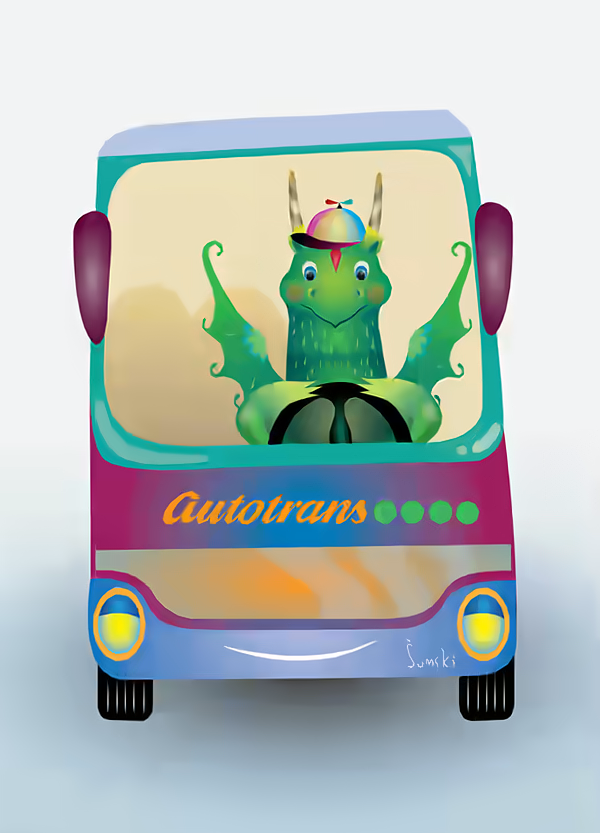}
    \captionsetup{justification=centering}
    \subcaption*{VVC\\33.3/0.069}
  \end{minipage}
  \begin{minipage}{0.32\hsize}
    \centering
    \includegraphics[width=\hsize]{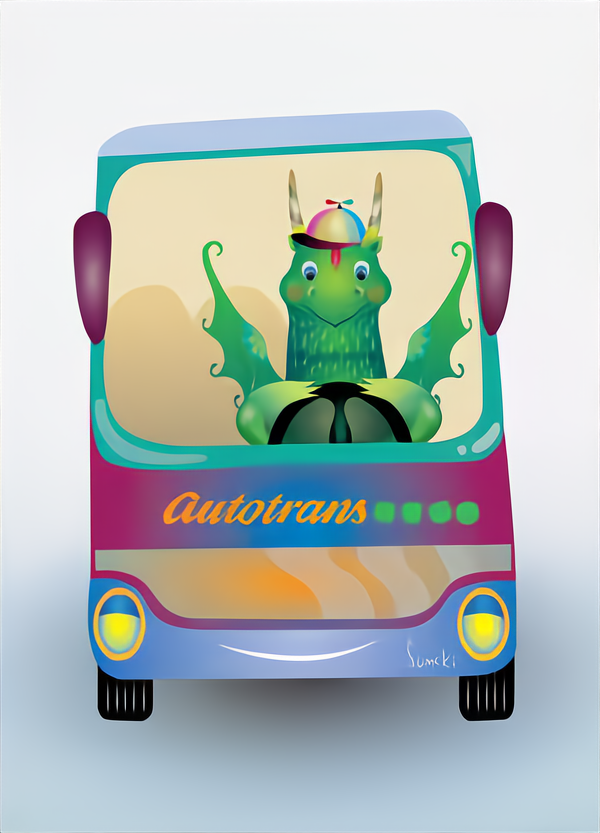}
    \captionsetup{justification=centering}
    \subcaption*{WACNN~\cite{ZouCVPR22}\\32.4/0.064}
  \end{minipage}
  \caption{Qualitative comparison with other compression methods on a vector art.}
  \label{fig:qua_vector}
\end{figure*}

\begin{figure}[t]
  \centering
  \begin{minipage}{0.32\hsize}
    \centering
    \includegraphics[width=\hsize]{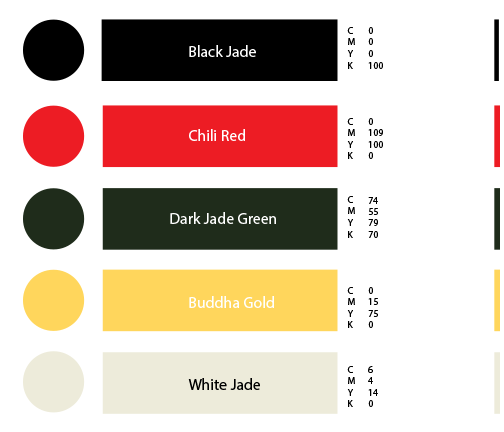}
    \captionsetup{justification=centering}
    \subcaption*{Input\\PSNR/BPP}
  \end{minipage}
  \begin{minipage}{0.32\hsize}
    \centering
    \includegraphics[width=\hsize]{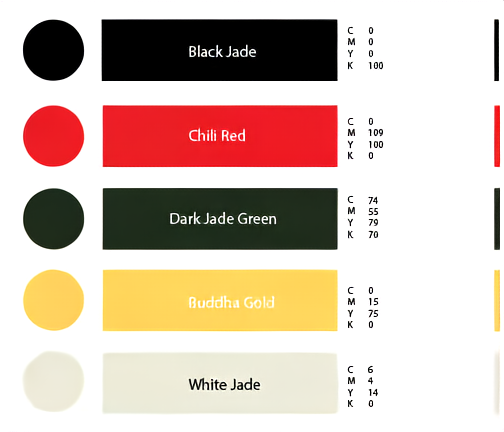}
    \captionsetup{justification=centering}
    \subcaption*{Ours\\35.4/0.099}
  \end{minipage}
  \begin{minipage}{0.32\hsize}
    \centering
    \includegraphics[width=\hsize]{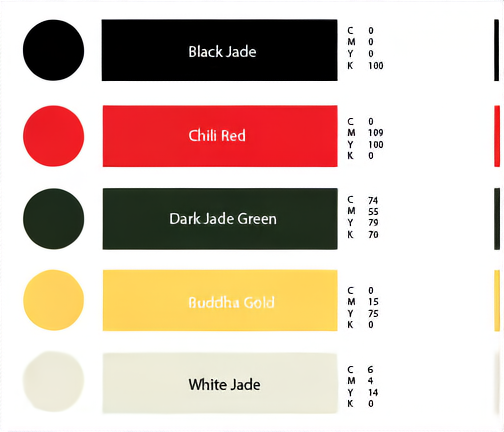}
    \captionsetup{justification=centering}
    \subcaption*{Without adapters\\34.5/0.090}
  \end{minipage}
  \caption{Qualitative results on the effectiveness of adapters on a vector art.}
  \label{fig:abl}
\end{figure}

The comparison results with other compression methods are shown in Figs.~\ref{fig:qua_natural}, \ref{fig:qua_comic}, \ref{fig:qua_pen}, and \ref{fig:qua_vector}.
The ablation results on the effectiveness of adapters are shown in Fig.~\ref{fig:abl}.

\end{document}